\documentclass{article}

\usepackage{arxiv}
\usepackage{amsmath}
\usepackage[utf8]{inputenc} 
\usepackage[T1]{fontenc}    
\usepackage{hyperref}       
\usepackage{url}            
\usepackage{booktabs}       
\usepackage{amsfonts}       
\usepackage{nicefrac}       
\usepackage{microtype}      
\usepackage{graphicx}
\usepackage{natbib}         
\usepackage{tabularx}
\usepackage{ragged2e}
\usepackage{array}
\usepackage{float}
\usepackage{placeins} 
\usepackage{enumitem} 

\graphicspath{ {figures/} }

\usepackage{scalerel}
\usepackage{tikz}
\usetikzlibrary{svg.path}
\definecolor{orcidlogocol}{HTML}{A6CE39}
\tikzset{
  orcidlogo/.pic={
    \fill[orcidlogocol] svg{M256,128c0,70.7-57.3,128-128,128C57.3,256,0,198.7,0,128C0,57.3,57.3,0,128,0C198.7,0,256,57.3,256,128z};
    \fill[white] svg{M86.3,186.2H70.9V79.1h15.4v107.1z M78.6,68.7c-5.6,0-10.1-4.5-10.1-10.1c0-5.6,4.5-10.1,10.1-10.1c5.6,0,10.1,4.5,10.1,10.1C88.7,64.2,84.2,68.7,78.6,68.7z M170.4,186.2h-15.4V121.1c0-16.2-5.5-27.2-20.3-27.2c-14.9,0-20.8,11-20.8,27.2v65.1H98.4V79.1h15.4v13.5c4.5-8.3,16.1-16.5,30.8-16.5c22.9,0,39.6,18.9,39.6,42.2V186.2z};
  }
}
\newcommand\orcidicon[1]{\href{https://orcid.org/#1}{\mbox{\scalerel*{
\begin{tikzpicture}[yscale=-1,transform shape]
\pic{orcidlogo};
\end{tikzpicture}
}{|}}}}

\title{The Adaptation Paradox: Agency vs. Mimicry in Companion Chatbots}

\author{
  T. James Brandt\thanks{Both authors contributed equally to this research.} \orcidicon{0009-0000-8294-6235} \\
  University of Minnesota\\
  1985 Buford Ave\\
  St. Paul, MN 55108, USA \\
  \texttt{bran1400@umn.edu} \\
  \And
  Cecilia Xi Wang\footnotemark[1] \orcidicon{0000-0003-4618-8849} \\
  University of Minnesota\\
  1985 Buford Ave\\
  St. Paul, MN 55108, USA \\
  \texttt{ceciw@umn.edu} \\
}

\begin{document}
\maketitle

\begin{abstract}
Generative AI powers a growing wave of companion chatbots, yet principles for fostering genuine connection remain unsettled. We test two routes: visible user authorship versus covert language-style mimicry. In a preregistered $3 \times 2$ experiment ($N = 162$), we manipulated user-controlled avatar generation (none, premade, user-generated) and Language Style Matching (LSM) (static vs.\ adaptive). Generating an avatar boosted rapport ($\omega^2 = .040, p = .013$), whereas adaptive LSM underperformed static style on personalization and satisfaction ($d = 0.35, p = .009$) and was paradoxically judged less adaptive ($t = 3.07, p = .003, d = 0.48$). We term this an \textit{Adaptation Paradox}: synchrony erodes connection when perceived as incoherent, destabilizing persona. To explain, we propose a stability-and-legibility account: visible authorship fosters natural interaction, while covert mimicry risks incoherence. Our findings suggest designers should prioritize legible, user-driven personalization and limit stylistic shifts rather than rely on opaque mimicry.
\end{abstract}


\section{Introduction}

\subsection{Personalizing AI Companions: Two Paths to Connection}
Artificial intelligence is increasingly entering the terrain of long-term, relational interaction. Beyond task assistance, contemporary conversational agents are positioned as companions, coaches, and confidants—systems meant to feel present, attentive, and responsive \citep{chandra2025, skjuve2023, liu2024, koh2025, polyportis2025}. As this shift accelerates, a central design question emerges, particularly as these interactions often manifest as parasocial relationships: one-sided, intimate connections that users form with media figures and, increasingly, with AI \citep{horton1956, kherraz2024, chandra2025, liu2024, pentina2023, zhang2025}. What actually builds rapport with a conversational agent—giving users visible, creative control over its presentation, or engineering it to invisibly adapt its linguistic style to the user? The promise of the latter is seductive: if an AI mirrors our phrasing, it should feel more “like us” and thus more personal, tapping into fundamental motivations for social connection and effectance—the desire to understand and predict other agents \citep{epley2007}. Yet an alternative hypothesis is equally plausible: user-visible, creative agency may matter more because it fosters ownership and a coherent mental model of the agent \citep{schlimbach2023}.

This tension is especially salient as the benefits of anthropomorphic design are not always straightforward. Both meta-analyses and systematic reviews show that while humanlike features are common design choices, their impact on user outcomes is highly context-dependent and often inconsistent \citep{blut2021, oh2021, klein2025}. For example, the interplay between visual, identity, and conversational cues can lead to complex interaction effects, where adding more cues does not always improve perceptions of humanness in a linear fashion \citep{go2019, chen2024, tsai2024}. Some studies find that anthropomorphic visual cues can even be detrimental, reducing self-disclosure in sensitive contexts like counseling \citep{kang2024, lucas2014}, while others report that visual cues alone are insufficient to improve user evaluations without complementary conversational strategies \citep{tsai2024}. While humanlike features are often assumed to increase user acceptance, recent studies show that users do not always prefer them \citep{aumuller2024}, suggesting that the quality and agency involved in personalization may be more critical than mere human-likeness.

\subsection{Theoretical Tensions: User Authorship versus System Adaptivity}
Extant literature offers foundational theories but provides divergent predictions. The CASA/MAIN perspective suggests that humanlike cues—visual embodiment and linguistic alignment—readily elicit social responses \citep{nass2000, sundar2008}. Research on co-creation indicates that even lightweight acts of customization can elevate identification and perceived value through psychological ownership \citep{norton2012, ryan2000}. In parallel, work on Language Style Matching (LSM) shows that subconscious linguistic synchrony correlates with rapport in human dyads \citep{ireland2011, niederhoffer2002}. Together, these strands imply two paths to connection: user-directed, visible adaptation (e.g., avatar creation) and system-driven, invisible adaptation (e.g., algorithmic LSM).

Translating these mechanisms into generative-AI chat, however, introduces a critical challenge: invisible mimicry can be technically correct yet perceptually illegible. If style shifts are fast or inconsistent, users may experience the agent as unstable or incoherent. This risk—that imperceptible adaptation can violate user expectations for coherence—is a central concern in human-AI interaction design \citep{amershi2019}. The upshot is a live design dilemma: should we invest in perfecting covert mimicry, or in empowering visible, legible control?

For clarity, we use \textbf{adaptability} to mean user-initiated control (authorship, agency) and \textbf{adaptivity} to mean system-initiated mimicry, following HCI usage \citep{schlimbach2023}.

\subsection{The Current Study and Contributions}
This paper addresses that dilemma with an empirical test. We report on a preregistered 3 × 2 between-subjects experiment in which 162 U.S. adults recruited via Prolific engaged in a 10-minute chat with a companion chatbot. We orthogonally manipulated two factors: Avatar Type, varying the degree of user-visible creative agency (None vs. Premade selection vs. Generated creation), and LSM, varying whether the chatbot’s style was held Static or adapted turn-by-turn. Our confirmatory outcomes focused on perceived personalization, social presence, trust, rapport, engagement, and satisfaction. Critically, this paper treats the underlying system as an experimental apparatus; a detailed treatment of its architecture is outside the scope of this work, which focuses on the human-subjects evaluation.

Three findings frame the contribution. First, granting users creative agency through avatar generation significantly improved rapport ($\omega^2$ = .040, p = .013). Second, in a pattern we term the "Adaptation Paradox," participants rated the static style as significantly more personal and satisfying (p = .009 for satisfaction) than the adaptive one, despite objective linguistic analysis showing the adaptive algorithm better maintained synchrony over time. Third, exploratory analyses suggest the adaptive pipeline’s variability reduced perceived coherence, whereas avatar creation nudged conversations toward more personal content. We call this pattern the Adaptation Paradox: an AI can be objectively more adaptive yet feel less personal. The paradox highlights a design principle for companion systems: in relational contexts, adaptation must be not only accurate but legible, stable, and attributable to an intelligible persona. Mirroring users’ style without a consistent narrative can degrade the very experience it aims to improve.

This work makes three contributions. (1) Empirically, we provide preregistered evidence from a controlled 3 × 2 study that user-visible agency (avatar generation) boosts relational outcomes, while invisible LSM underperforms a static baseline. (2) Mechanistically, we show a divergence between objective synchrony and subjective adaptiveness, linking this gap to perceptual legibility—a nuance often overlooked when importing findings from human-human LSM into HCI. (3) For design and ethics, we distill actionable guidance grounded in established principles for human-AI interaction \citep{amershi2019}: prioritize user-directed, visible adaptation, and ensure system-driven adaptation is perceptible and coherent.

\begin{figure}[htbp]
  \centering
  \includegraphics[width=\linewidth]{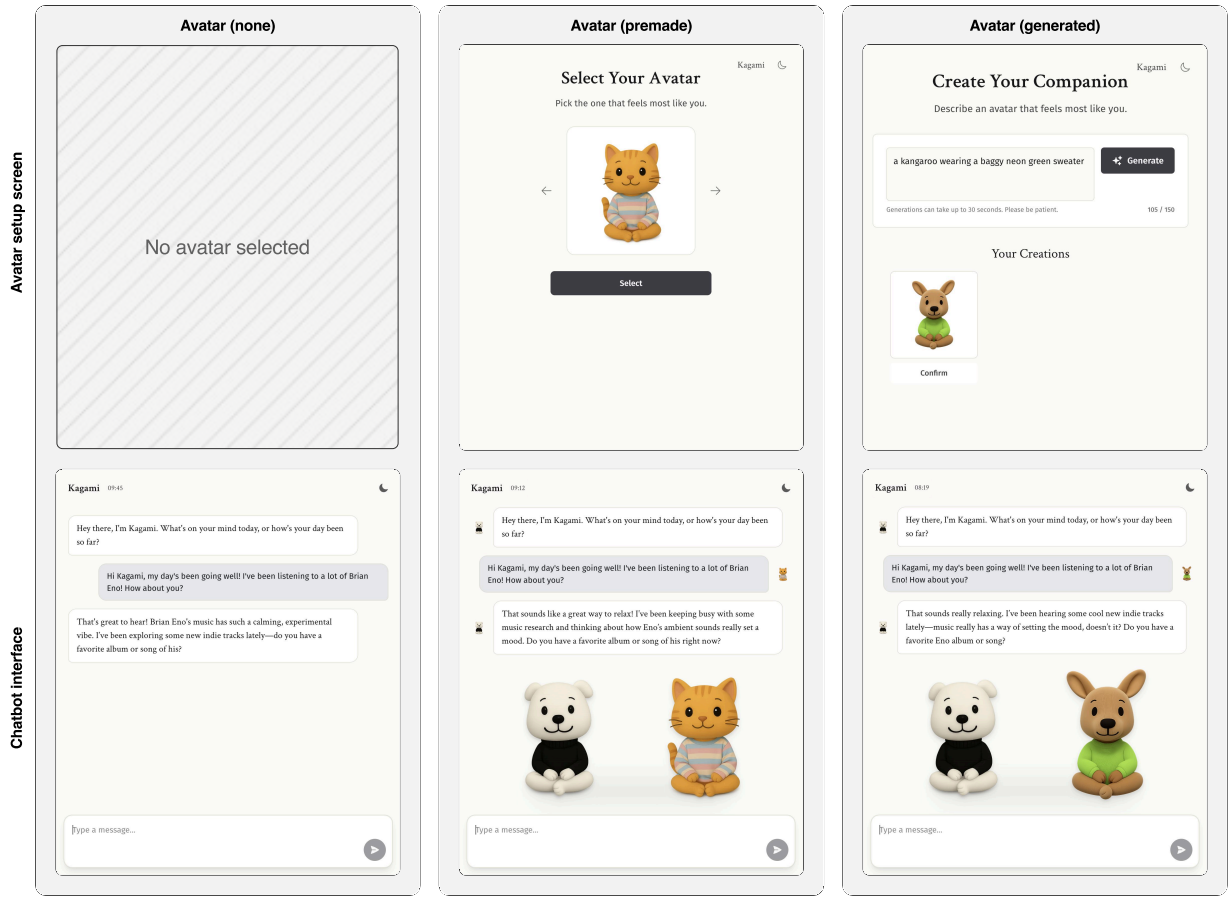}
  \caption{Visual overview of the user flow and interfaces for the three `Avatar Type` experimental conditions: (left) None, (middle) Premade, and (right) Generated. The top row displays the avatar setup screen for each condition, illustrating the different levels of user agency: no interaction, selection from a gallery, or creation via a text prompt. The bottom row shows the resulting chatbot interface, demonstrating how the user's choice (or lack thereof) was visually represented in the conversation.}
  \label{fig:exp_overview}
\end{figure}
\FloatBarrier

\section{Research Questions \& Hypotheses}

\subsection{Research Questions}
\begin{description}
    \item[RQ1:] Does granting user-visible creative agency through avatar creation improve relational outcomes compared to premade avatars or no avatar?
    \item[RQ2:] Does an adaptive linguistic style improve perceived personalization and satisfaction relative to a static style?
    \item[RQ3:] When outcomes diverge, what mechanisms (e.g., objective–subjective gaps, content shifts, stability/legibility) best explain the Adaptation Paradox?
\end{description}

\subsection{Confirmatory Hypotheses (Preregistered)}
\begin{description}
    \item[H1:] The presence of a user-chosen avatar (either Premade or Generated) will increase positive relational outcomes compared to a no-avatar control. This hypothesis draws from theories of Anthropomorphism \citep{blut2021} and the Proteus Effect \citep{yee2007}.
    \item[H2:] An adaptive linguistic style will increase positive relational outcomes compared to a static style.
    \item[H3 (Interaction):] The combination of a user-generated avatar and an adaptive linguistic style will produce the strongest positive relational outcomes, creating a synergistic effect. This is based on the principle of congruence between visual and conversational cues \citep{go2019}.
\end{description}
\FloatBarrier

\section{Background and Theoretical Framework}

\subsection{Anthropomorphism and the Psychology of Relational AI}
Conversational agents can evoke social responses because humans have a natural tendency to anthropomorphize non-human agents—attributing to them humanlike characteristics, motivations, and intentions. Foundational theory in social cognition posits this is not an arbitrary tendency but is driven by three key psychological determinants: \textbf{effectance motivation}, the need to explain and predict an agent’s behavior to interact with it effectively; \textbf{sociality motivation}, the fundamental human need for social connection; and \textbf{elicited agent knowledge}, where knowledge about humans serves as the most accessible cognitive model for understanding an unknown entity \citep{epley2007}.

This framework explains why both visual and linguistic human-likeness are pursued in relational AI design. Visual embodiment (an avatar) and linguistic alignment (mimicking style) are powerful cues that can make an agent feel more like a coherent, understandable, and socially present partner, aligning with the principles of the Computers Are Social Actors (CASA) paradigm \citep{nass2000, gambino2020, klein2025}. The MAIN model further specifies this process, arguing that distinct technological affordances on an interface act as heuristic cues, which users automatically interpret to make judgments about an agent's character and credibility \citep{sundar2008}. Our study investigates two distinct methods of leveraging these principles: empowering user agency in visual presentation versus engineering covert linguistic mimicry.

\subsection{The Power of Creative Agency in Visual Embodiment}
The first route to fostering connection is through \textbf{adaptability}: giving users direct control to customize the agent's appearance. Granting users this creative agency, even in modest doses, can strengthen identification and attachment. This is explained by several converging psychological mechanisms:

\begin{itemize}
    \item \textbf{Psychological Ownership and the IKEA Effect}: People value things more when they invest their own labor in creating them. The act of designing an avatar fosters a sense of ownership (``this is mine'') that deepens the user's connection to the agent \citep{norton2012, pierce2001, dawkins2017, vandyne2004}.
    
    \item \textbf{Self-Determination Theory (SDT)}: The process of creating an avatar supports innate psychological needs for \textbf{autonomy} (feeling volitional) and \textbf{competence} (feeling effective), which in turn fosters deeper engagement and positive affect \citep{ryan2000, bandura1977}.
    
    \item \textbf{The Proteus Effect}: Beyond simple preference, a user's behavior begins to conform to the characteristics of their avatar. For instance, users with avatars they perceive as attractive may behave more confidently and sociably in virtual environments \citep{yee2007, rheu2022}. By allowing users to generate their own avatar, the system provides a visual anchor for identity that can directly shape the user's interaction style, making them more likely to engage in the kind of personal disclosure that builds rapport.
\end{itemize}

Together, these theories predict that giving users creative agency over their chatbot's avatar is not mere decoration; it is a powerful act of co-creation that should increase social presence, rapport, and satisfaction. Indeed, this has been demonstrated empirically. Experimental work in gaming has shown that the act of customization itself has a greater impact on player-avatar identification than objective appearance similarity, underscoring the psychological importance of authorship \citep{wang2022}. More recently, studies on conversational agents have found that personalizing a chatbot's avatar and personality can significantly increase therapeutic bond \citep{vossen2024}, while the act of customizing an avatar to reflect a chosen archetype can directly influence subsequent moral behavior \citep{pena2024}. This effect is theoretically grounded in the CASA paradigm and theories of objective self-awareness, which posit that embodying a personalized avatar makes the interaction more salient and self-relevant \citep{koek2025}. While the link between customization and outcomes can be complex \citep{paul2024}, the evidence points to user agency as a powerful design lever.

\subsection{Linguistic Alignment and the Adaptation Paradox}
In human-human interaction, people nonconsciously align their use of function words (e.g., pronouns, articles), a phenomenon rooted in Communication Accommodation Theory \citep{giles1991} and computationally operationalized as LSM \citep{gonzales2010}. LSM focuses on these ``style words'' because they are used subconsciously and are believed to be a powerful signal of psychological states, reflecting \textit{how} people are communicating rather than simply \textit{what} they are communicating about \citep{tausczik2010, pennebaker2003}. Consequently, higher LSM is a robust correlate of positive social outcomes, including increased rapport, mutual understanding, group cohesiveness, and relationship stability, making it a key mechanism for building social connection \citep{gonzales2010, ireland2011, niederhoffer2002}. The assumption in HCI has been that if a chatbot dynamically mimics a user’s linguistic style, it will feel more attuned and personal \citep{nye2014, d2013, gnewuch2020, van2020}. Indeed, studies have shown that chatbot alignment can decrease perceived task workload \citep{spillner2021} and increase perceived trustworthiness, particularly for users with a more considerate, less-involved conversational style \citep{hoegen2019, colton2024}. However, these effects are not universal, as the overall positive impact of human-like cues is often small and highly moderated by context \citep{klein2025}. The mechanism for this effect on trust, however, may not be direct; rather, anthropomorphic and interactive features appear to build trust by first enhancing the perceived quality of the communication itself \citep{nguyen2023}.

However, translating this mechanism to generative AI introduces two significant risks that can lead to an \textbf{Adaptation Paradox}, where an objectively more adaptive system feels less personal:

\begin{enumerate}
    \item \textbf{Legibility}: If the AI's adaptation is entirely covert, users may not notice it or, if they do, may not attribute it to a coherent personality. Instead of perceiving care, they may perceive randomness or instability, undermining the agent's credibility and the user's trust \citep{sundar2008, amershi2019, lim2009}. This aligns with findings that when an agent's identity cues (e.g., an "expert" label) are incongruent with its conversational behavior (e.g., generic feedback), users form negative perceptions due to the expectancy violation \citep{go2019, janson2023, white2021, rheu2024}. A lack of system intelligibility is a well-documented barrier to user trust, satisfaction, and adoption of intelligent systems; providing explanations for system behavior is a core principle for helping users form accurate mental models and correct system errors \citep{lim2009, kulesza2015}.
    
    \item \textbf{Stability}: High turn-to-turn stylistic variability, even if it maintains high average LSM, can erode the sense of a consistent persona. This violates users' expectations for predictability in a conversational partner, which is a foundational element for reducing uncertainty and building trust in initial interactions \citep{berger1975, nass2000, lee2004, pennebaker1999}. Indeed, recent work on dynamic personality infusion suggests that a \textit{consistent} and stable persona is a key driver of positive outcomes like trust and enjoyment \citep{kovacevic2024}. From a sociolinguistic perspective, effective communication requires adhering to a linguistic register appropriate for the agent's social role and context; a style that is incoherent or contextually inappropriate can reduce satisfaction and trust \citep{chaves2022, klein2025, chen2024}. Recent work confirms that while LLM agents can align linguistically, this often comes at the cost of persona consistency, particularly when the agent's persona conflicts with gender or role-based stereotypes held by the user \citep{frisch2024, schillaci2024}.
\end{enumerate}

This creates a critical gap between objective synchrony (which an algorithm can be tuned to achieve) and ``felt'' adaptation (which depends on the user's ability to form a stable mental model of the agent).

This tension can be framed by Expectancy Violation Theory (EVT) \citep{white2021}, which posits that individuals hold expectations about a partner's behavior and react negatively when those expectations are violated. While theories of mimicry predict a positive effect from an adaptive style, EVT suggests a potential downside. Users may hold a strong expectation for a \textit{stable and coherent persona} \citep{nass2000}. An agent with high turn-to-turn stylistic variability, even if technically synchronous, risks violating this expectation of coherence. Such a violation could lead to a negative evaluation, making the agent feel less credible and, paradoxically, less personal. Our experiment is therefore designed to adjudicate this tension directly.

\subsection{Theoretical Synthesis: Parasocial Bonds and Testable Hypotheses}
The bonds users form with companion chatbots are best understood as a modern form of \textbf{parasocial relationship (PSR)}—a one-sided, mediated connection that feels intimate to the user \citep{horton1956}. A major meta-analysis synthesizing over 100 studies confirms that while anthropomorphism generally has a positive impact on users' intention to use a service AI, its effects are highly context-dependent and are mediated by relational outcomes like rapport and satisfaction \citep{blut2021}. This underscores that not all humanlike cues are equally effective. Qualitative studies on chatbots like Replika further confirm that users form deep, emotionally significant bonds, navigating a shared history with an entity they know is not human \citep{skjuve2023, pentina2023}.

This framework sharpens our central research question: In fostering a parasocial bond, what matters more—the visible authorship of \textit{adaptability} or the invisible mimicry of \textit{adaptivity}? Our study is designed to place these two mechanisms in direct empirical tension by testing the competing predictions of established theories. On one hand, theories of anthropomorphism and mimicry suggest that both visual embodiment and linguistic alignment should enhance relational outcomes. On the other hand, principles of psychological ownership and expectancy violation suggest that the \textit{quality} of these cues is paramount: user-directed authorship may be more powerful than mere presence, and a stable, coherent persona may paradoxically feel more personal than a technically adaptive but perceptually inconsistent one. Our experiment is designed to adjudicate these tensions directly.

\begin{figure}[htbp]
  \centering
  \includegraphics[width=\linewidth]{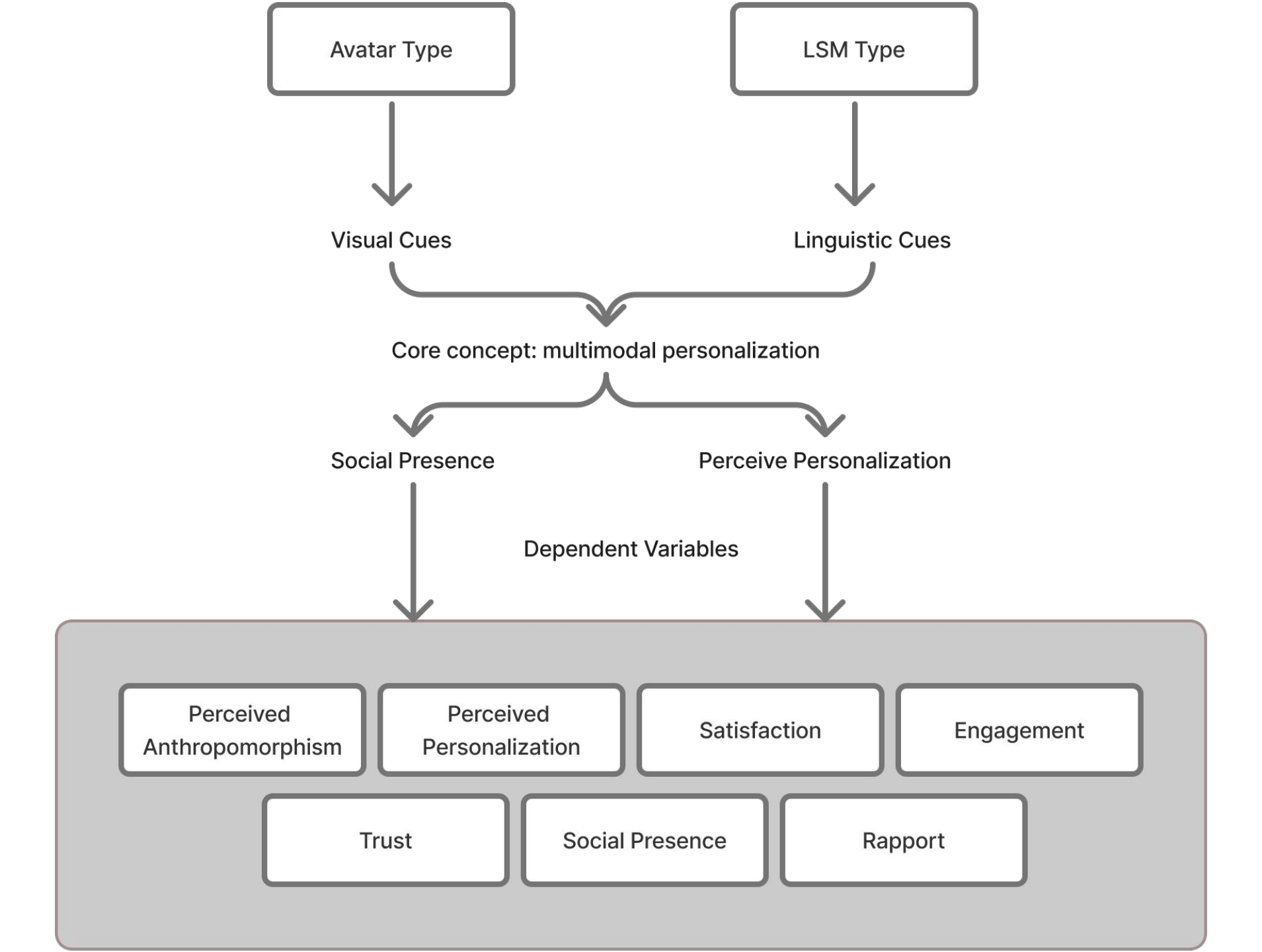}
  \caption{Conceptual model of the hypothesized effects. We test two routes to personalization: user-visible agency (Avatar Type) and system-driven mimicry (LSM Style), examining their impact on relational outcomes.}
  \label{fig:conceptual_model}
\end{figure}
\FloatBarrier

\section{Methods}
We conducted a preregistered 3 (Avatar Type: None vs. Premade vs. Generated) × 2 (Linguistic Style: Static vs. Adaptive) between-subjects experiment to examine the effects of user agency and linguistic mimicry on relational outcomes with a companion chatbot. The study design, hypotheses, and analysis plan were preregistered on the Open Science Framework (\url{https://osf.io/f4h5b}).

\subsection{Participants and Recruitment}
\begin{figure}[htbp]
  \centering
  \includegraphics[width=0.8\linewidth]{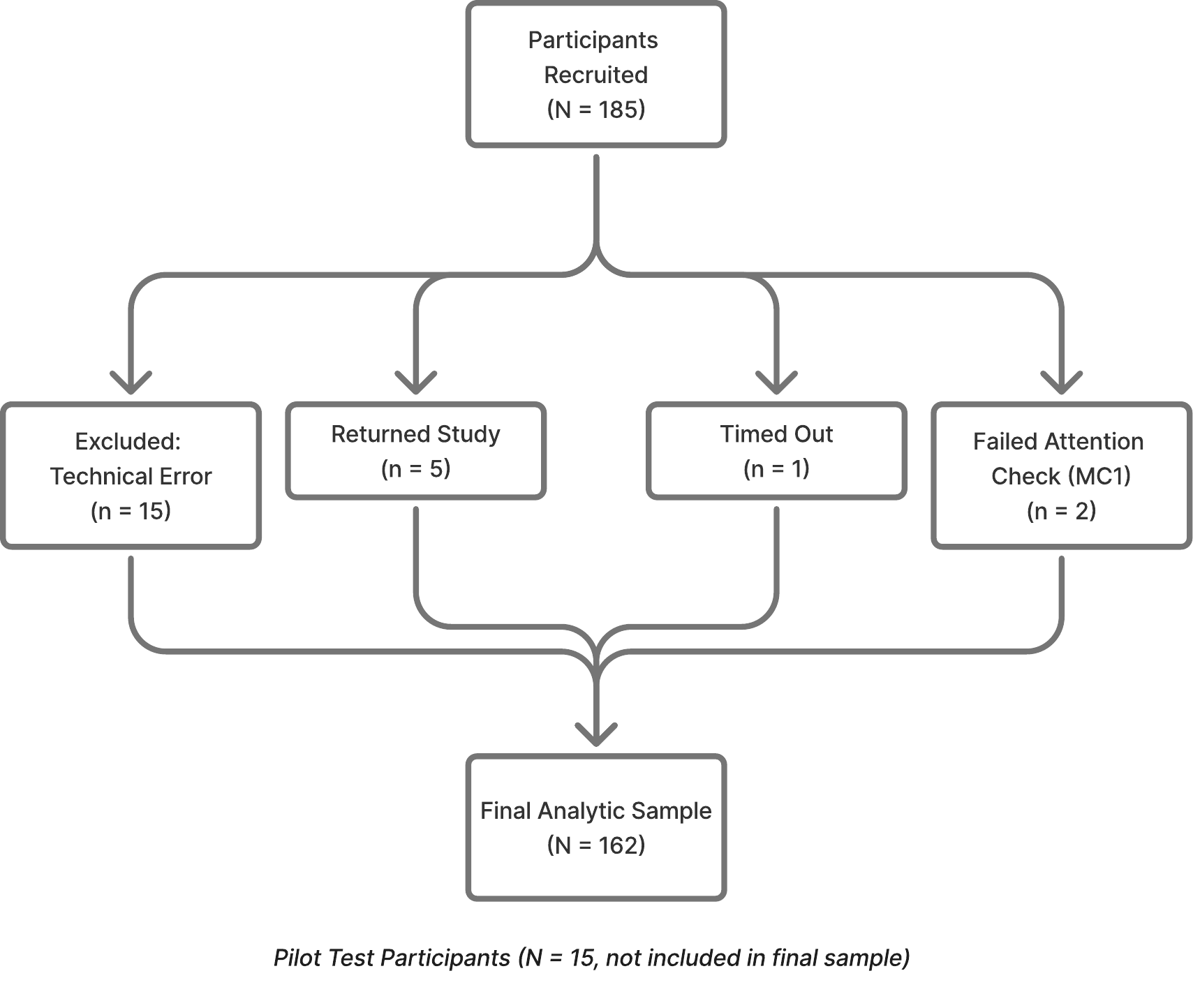}
  \caption{CONSORT diagram detailing the flow of participants from recruitment through exclusion to the final analyzed sample ($N = 162$). \textit{Note: A 15-participant pilot sample (not shown) was collected prior to the main study and excluded here. Final participants were randomized evenly across the six experimental conditions.}}
  \label{fig:consort}
\end{figure}

The study received IRB approval from the University of Minnesota (STUDY00025677). Between June 28 and June 29, 2025, we recruited 185 participants for the main study from the United States via the Prolific online research platform (\url{www.prolific.com}) \citep{prolific2014} (excluding a 15-participant pilot). Following preregistered exclusion criteria, 23 participants were excluded for the following reasons: technical errors preventing chat log generation (n=15), voluntarily returning the study (n=5), failing the avatar awareness manipulation check (n=2), or timing out before completion (n=1). This yielded a final analytic sample of 162 participants. A full CONSORT flow diagram is provided in Figure \ref{fig:consort} \citep{moher2010}.

Participants were required to be 18 or older, fluent in English, and use a desktop or laptop computer. Random assignment resulted in balanced conditions, with approximately 27 participants per cell. The 15–20 minute session was compensated with \$3.50. An a priori power analysis using G*Power 3.1 \citep{faul2007} indicated that a sample of 158 was needed to detect a medium effect size (f = 0.25) with 80\% power at $\alpha = .05$; our final sample of 162 was therefore adequately powered. Participant demographics are summarized in Table \ref{tab:demographics}.

\begin{table}[htbp]
  \centering
  \caption{Participant Demographics ($N = 162$). \textit{Note: Percentages may not sum to 100 due to rounding.}}
  \label{tab:demographics}
  \begin{tabular}{l l r}
    \toprule
    \textbf{Characteristic} & \textbf{Category} & \textbf{N (\%)} \\
    \midrule
    \textbf{Age} & Mean (SD) & 41.32 (13.38) \\
    \midrule
    \textbf{Gender Identity} & Man & 86 (53.1\%) \\
                         & Woman & 75 (46.3\%) \\
                         & Non-binary / Other & 1 (0.5\%) \\
    \midrule
    \textbf{Race/Ethnicity} & White & 111 (68.5\%) \\
                            & Black or African American & 23 (14.2\%) \\
                            & Multiracial or Other & 13 (8.0\%) \\
                            & Asian & 8 (4.9\%) \\
                            & Hispanic or Latino & 5 (3.1\%) \\
                            & Other & 2 (1.2\%) \\
    \bottomrule
  \end{tabular}
\end{table}

\subsection{Apparatus and Manipulations}
The experiment was conducted on a custom-built, single-page web application named Kagami. The system is treated as an experimental apparatus; a detailed treatment of the underlying model (OpenAI GPT-4.1-nano \citep{openai2025gpt41nano}) and prompting architecture is outside the scope of this work, which focuses on the human-subjects evaluation. We manipulated two factors.

\subsubsection{Avatar Type (User Agency)}
This factor manipulated the degree of user-visible creative agency through three levels:
\begin{itemize}
    \item \textbf{None}: Participants proceeded directly to the chat interface with no avatar displayed for the chatbot.
    \item \textbf{Premade}: Participants selected one of seven predefined non-human animal avatars from a gallery to represent the chatbot.

\begin{figure}[htbp]
  \centering
  \includegraphics[width=0.8\linewidth]{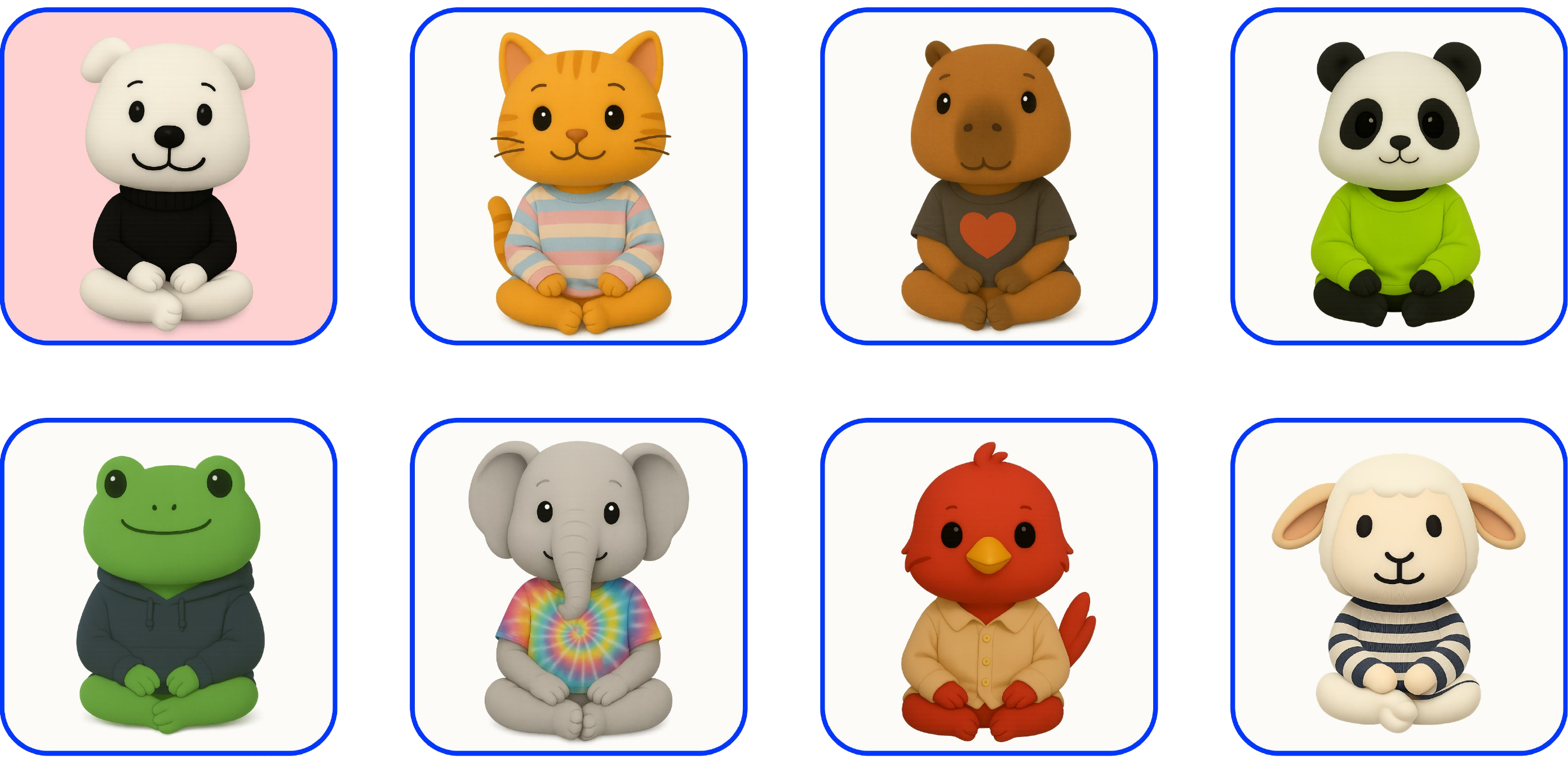}
  \caption{The seven stylistically-matched premade avatars available for selection in the \textit{Premade} condition. The top-left avatar (white dog) was not selectable by participants; it served only as the visual style template for the \textit{Generated} condition to maintain stylistic consistency and as the default visual persona of the chatbot in the interface.}
  \label{fig:premade_avatars}
\end{figure}
    
    \item \textbf{Generated}: Participants authored a unique avatar by writing a short text description (e.g., "a wise old owl wearing reading glasses"), which was rendered by the OpenAI GPT-Image-1 generative image model \citep{openai2025gptimage1} in a consistent art style.
\end{itemize}
\FloatBarrier

\subsubsection{Linguistic Style (System Adaptivity)}
This factor manipulated the chatbot's conversational style. Participants were blind to this condition.
\begin{itemize}
    \item \textbf{Static}: The chatbot used a consistent, non-adaptive linguistic style throughout the conversation, guided by a single system prompt instructing it to be "helpful, friendly, and moderately formal."
    \item \textbf{Adaptive}: A multi-stage backend pipeline dynamically adapted the chatbot's style. For each turn, the pipeline analyzed the last five user utterances using key LIWC-22 function word categories (e.g., pronouns, articles, prepositions) \citep{boyd2022}. It then translated the user's aggregate style profile into a natural language instruction that was prepended to the system prompt. For example, if a user's style was detected as informal and social, the system might generate the instruction, ``Adopt a casual, friendly, and social tone in your response,'' to guide the LLM's next turn. This approach is conceptually aligned with recent advances in activation engineering, which enable dynamic, inference-time steering of LLM behavior without model retraining \citep{wang2024}.
\end{itemize}

\subsection{Procedure}
After providing informed consent via a Qualtrics survey, participants completed a brief pre-survey containing demographic and baseline measures. They were then randomly assigned to one of six experimental conditions using Qualtrics' block randomization feature. Following assignment, participants were redirected to a custom web application where the experimental manipulation was delivered. Those in the \textit{Premade} or \textit{Generated} conditions first completed the brief avatar selection or creation step, after which all participants proceeded to a 10-minute, open-ended conversation with the chatbot. Upon completion, participants were automatically redirected back to Qualtrics for a post-interaction survey containing the outcome measures, manipulation checks, and open-ended feedback questions.

\subsection{Measures}
All primary outcomes were measured using validated multi-item scales with 5-point Likert-style responses (1 = \textit{Strongly Disagree}, 5 = \textit{Strongly Agree}). Composite scores were created by averaging items for each scale.
\begin{description}
    \item[Primary Outcomes] included seven preregistered dependent variables: Perceived Anthropomorphism \citep{blut2021}, Perceived Personalization \citep{paul2024}, Social Presence \citep{biocca2003}, Trust \citep{mayer1995}, Rapport \citep{colton2024}, Engagement \citep{hoffman2023}, and Satisfaction \citep{blut2021}.
    
    \item[Manipulation Checks] assessed awareness of the avatar condition and participants' subjective ratings of the chatbot's adaptiveness and consistency (e.g., "The chatbot's language style seemed to adapt to the way I was talking.").
    
    \item[Objective Linguistic and Content Measures] were derived from chat logs. Objective LSM was calculated turn-by-turn using a formula adapted from \citet{ireland2011}, quantifying similarity across nine function-word categories. Conversation Content Profiles were derived using BERTopic \citep{grootendorst2022} to thematically categorize conversations (e.g., PERSONAL, SUPERFICIAL, META-AI).
\end{description}

\subsection{Data Analysis}
Our analysis plan followed the preregistration. All statistical tests were two-tailed with an alpha level of .05, conducted in Python using the \texttt{statsmodels} \citep{seabold2010} and \texttt{pingouin} \citep{Vallat2018} libraries. Each of the seven primary outcomes was analyzed using a 3 (Avatar Type) × 2 (Linguistic Style) between-subjects ANOVA. We report omnibus F-tests with $\omega^2$ as our measure of effect size. Significant effects were followed by preregistered planned contrasts, for which we report Cohen’s d and 95\% confidence intervals (CIs).
\FloatBarrier

\section{Results}
\noindent\textbf{The study's central finding was a significant "Adaptation Paradox."} Contrary to the manipulation's intent, a t-test on perceived adaptiveness revealed that participants in the \textbf{Static} condition ($M = 4.14, SD = 0.83$) rated the chatbot as significantly more adaptive than those in the \textbf{Adaptive} condition ($M = 3.74, SD = 1.01$), $t(143.71) = 3.07, p = .003, d = 0.48$. This critical divergence—where the objectively more adaptive system was perceived as less so—is a core finding, not a failed manipulation check. The following sections unpack the consequences of this paradox on our primary outcomes and explore its underlying mechanisms.

\subsection{Manipulation Check for Avatar Type}
The Avatar Type manipulation was successful. Participants in the avatar conditions reported greater authorship over the chatbot's appearance than those in the no-avatar control, with the \textit{Generated} condition scoring significantly higher than the \textit{Premade} condition, confirming that our manipulation successfully varied the degree of perceived creative agency.

\begin{figure}[htbp]
  \centering
  \includegraphics[width=0.8\linewidth]{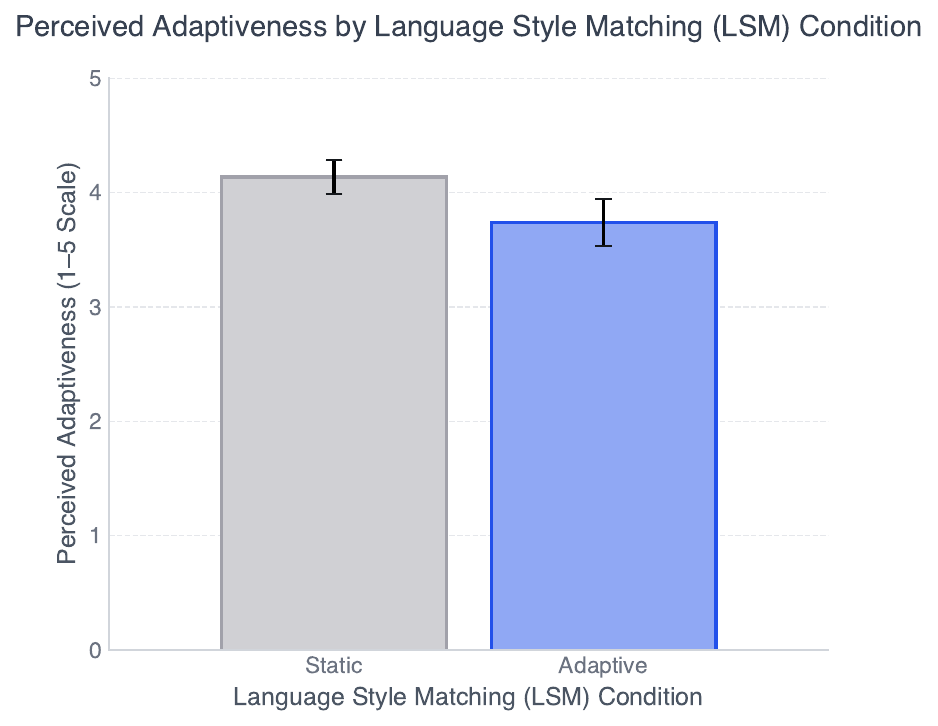}
  \caption{Perceived adaptiveness by linguistic style condition. Despite the Adaptive algorithm objectively maintaining synchrony better, participants rated the Static chatbot as significantly more adaptive, illustrating the core Adaptation Paradox.}
  \label{fig:perceived_adaptiveness}
\end{figure}

\subsection{Confirmatory Analysis of Primary Outcomes}
We analyzed each of our seven primary outcomes using a 3 (Avatar Type) × 2 (Linguistic Style) ANOVA. Full omnibus test results are reported in Table \ref{tab:anova_results}. All analyses were robust to violations of normality assumptions, as confirmed by preregistered non-parametric tests.
\FloatBarrier

\subsubsection{Main Effect of Avatar Type on Rapport}
Supporting H1, we found a significant main effect of Avatar Type on \textbf{Rapport}, $F(2, 156) = 4.49, p = .013, \omega^2 = .040$. However, a more nuanced picture emerged from preregistered post hoc tests, which revealed it is not the mere presence of an avatar but the user's \textbf{creative agency} in its generation that drives this effect. Specifically, there was no significant difference between the \textit{Generated} and \textit{None} conditions ($p_{\text{adj}} = .520$) or between the \textit{Premade} and \textit{None} conditions ($p_{\text{adj}} = .177$).

\begin{figure}[htbp]
  \centering
  \includegraphics[width=0.8\linewidth]{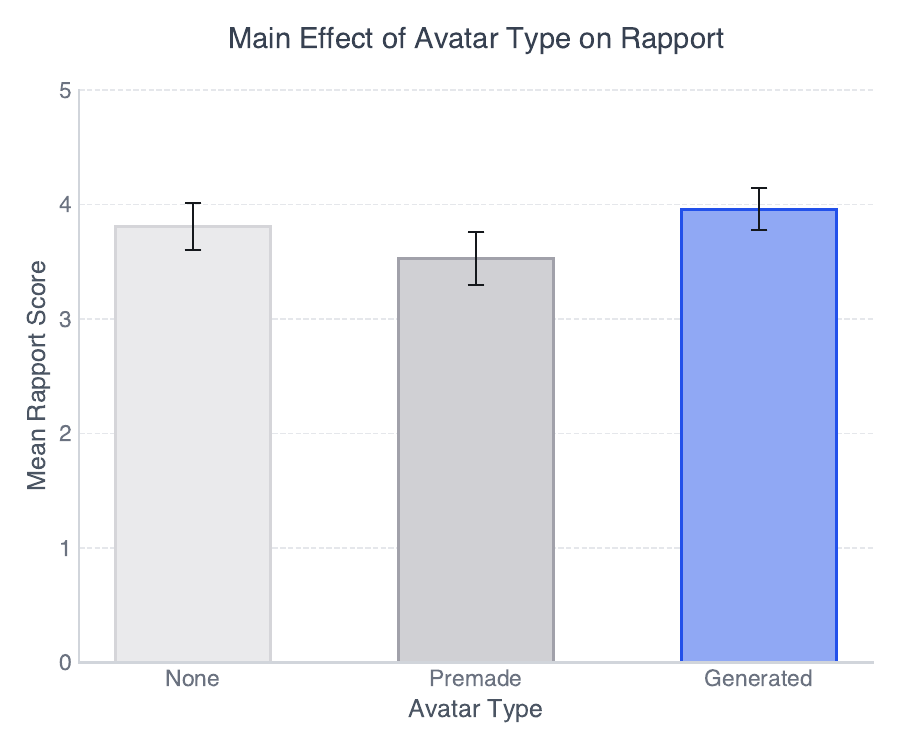}
  \caption{Main effect of Avatar Type on Rapport. The creative agency in the \textit{Generated} condition produced significantly higher rapport than selecting a \textit{Premade} avatar.}
  \label{fig:rapport_by_avatar}
\end{figure}

However, we found strong support for the importance of creative agency. The post hoc test showed that the \textbf{Generated} condition ($M = 3.96, SD = 0.82$) yielded significantly higher rapport than the \textbf{Premade} condition ($M = 3.53, SD = 0.89$), $p_{\text{adj}} = .013$, 95\% CI for difference $[0.07, 0.79]$. This result indicates that the act of co-creation, rather than mere avatar presence, was the key driver of this relational benefit (Figure \ref{fig:rapport_by_avatar}).

\FloatBarrier

\subsubsection{Main Effect of Linguistic Style}
In a direct contradiction of H2---and central to this paper's contribution---the \textit{Static} linguistic style consistently outperformed the \textit{Adaptive} style on key subjective outcomes. We found significant main effects of Linguistic Style, with the Static condition leading to higher ratings on \textbf{Perceived Anthropomorphism} ($F(1, 156) = 5.39, p = .022, \omega^2 = .027$), \textbf{Perceived Personalization} ($F(1, 156) = 4.06, p = .046, \omega^2 = .019$), and \textbf{Satisfaction} ($F(1, 156) = 7.00, p = .009, \omega^2 = .035$). This empirical pattern constitutes the \textbf{Adaptation Paradox}: a technically more adaptive system was perceived as less personal and satisfying.

\subsubsection{Interaction Effects}
The preregistered H3, which predicted a synergistic interaction between a generated avatar and an adaptive style, was not supported. Omnibus Avatar × Style interaction terms were not significant for any primary outcome (all $p > .05$). The benefits of user agency and the costs of invisible mimicry appeared to be largely additive.
\FloatBarrier

\begin{table*}[htbp]
  \centering
  \caption{Summary of 3 (Avatar Type) × 2 (Linguistic Style) ANOVA results for all primary outcomes ($N = 162$).}
  \label{tab:anova_results}
  \small 
  \setlength{\tabcolsep}{4pt}  
  \begin{tabular}{l rrr c rrr c rrr}
    \toprule
    & \multicolumn{3}{c}{\textbf{Avatar Type}} & \phantom{a} & \multicolumn{3}{c}{\textbf{Linguistic Style}} & \phantom{a} & \multicolumn{3}{c}{\textbf{Avatar × Style Interaction}} \\
    \cmidrule(lr){2-4} \cmidrule(lr){6-8} \cmidrule(lr){10-12}
    \textbf{Outcome} & $F(2, 156)$ & $p$ & $\omega^2$ && $F(1, 156)$ & $p$ & $\omega^2$ && $F(2, 156)$ & $p$ & $\omega^2$ \\
    \midrule
    Rapport                    & 4.49 & \textbf{.013} & .040 && 2.86 & .093 & .011 && 2.24 & .110 & .014 \\
    Satisfaction               & 2.01 & .138 & .012 && 7.00 & \textbf{.009} & .035 && 2.50 & .085 & .017 \\
    Perceived Personalization         & 0.83 & .438 & .000 && 4.06 & \textbf{.046} & .019 && 1.18 & .309 & .002 \\
    Perceived Anthropomorphism        & 0.76 & .470 & .000 && 5.39 & \textbf{.022} & .027 && 0.43 & .651 & .000 \\
    Social Presence            & 2.57 & .080 & .019 && 2.09 & .151 & .006 && 2.65 & .074 & .020 \\
    Trust                      & 2.45 & .089 & .017 && 3.70 & .056 & .016 && 1.95 & .145 & .011 \\
    Engagement                 & 2.92 & .057 & .022 && 3.28 & .072 & .013 && 3.01 & .052 & .023 \\
    \bottomrule
  \end{tabular}
\end{table*}


\subsection{Unpacking the Adaptation Paradox: Exploratory Mechanisms}
To understand the divergence between objective system behavior and subjective user experience, we conducted exploratory analyses on the chat logs.

\subsubsection{Objective Synchrony vs. Perceived Coherence}
\begin{figure}[htbp]
  \centering
  \includegraphics[width=0.8\linewidth]{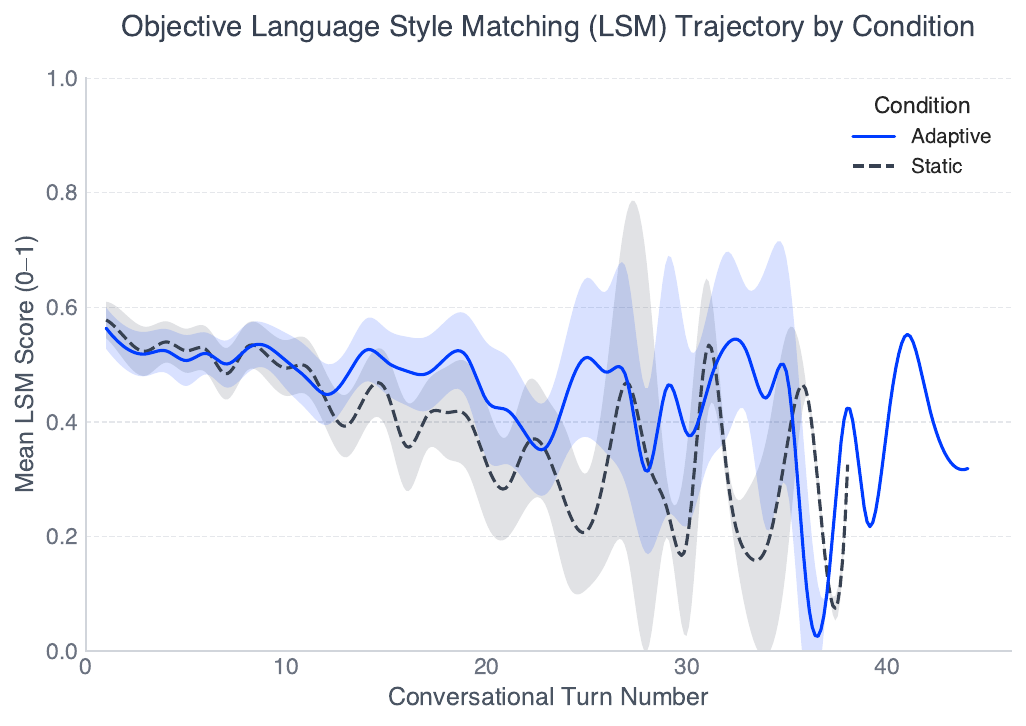}
  \caption{Objective LSM trajectory over conversation turns. The Adaptive algorithm successfully buffered against the natural decay in synchrony seen in the Static condition, yet was perceived as less adaptive.}
  \label{fig:lsm_trajectory}
\end{figure}

First, we confirmed the \textit{Adaptive} algorithm's technical efficacy. While there was no significant difference in average LSM between conditions, a linear mixed-effects model revealed a significant Condition × Turn interaction ($b = -0.006, p < .001$). This shows that LSM scores declined naturally over time in the \textit{Static} condition, whereas the \textit{Adaptive} algorithm successfully buffered against this decay, maintaining a more stable level of synchrony (Figure \ref{fig:lsm_trajectory}).

Despite this objective advantage, users perceived the \textit{Static} agent as more adaptive. Our exploratory analysis points to a powerful mechanism for this paradox: users do not perceive raw linguistic synchrony directly, but rather its emergent property—\textbf{perceptual coherence}. This data-driven hypothesis provides a clear direction for future studies, which could then experimentally test the causal role of perceptual stability in mediating user perceptions. Consistent with established Human-AI guidelines, our findings suggest that future work should prioritize ensuring perceptual coherence over simply maximizing objective synchrony \citep{amershi2019}.

\subsubsection{Content Shifts as a Behavioral Pathway for Avatar Benefits}
\begin{figure}[htbp]
  \centering
  \includegraphics[width=0.8\linewidth]{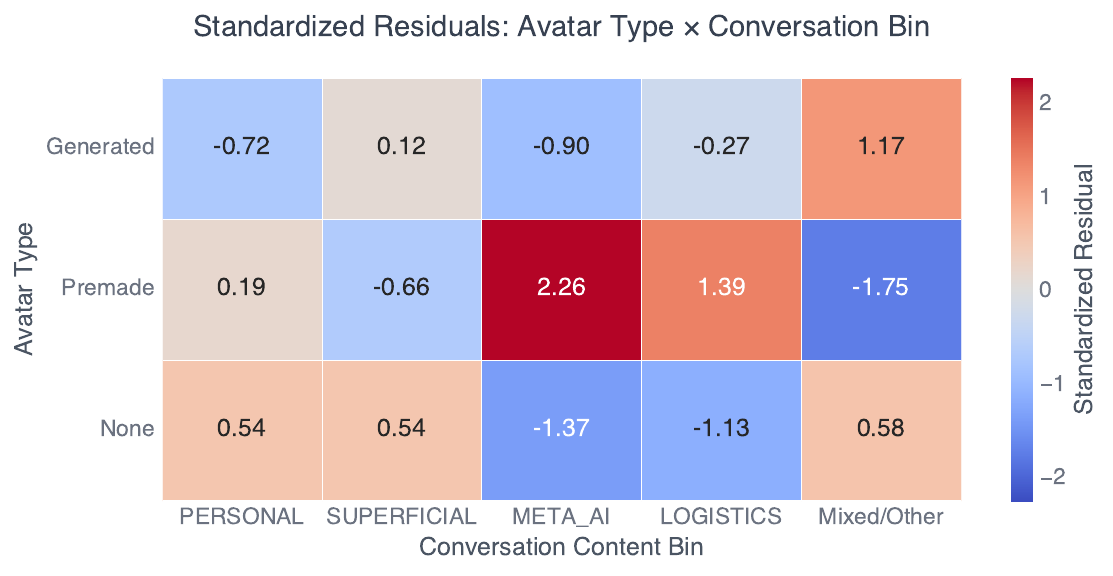}
  \caption{Standardized residuals for conversation content bins by avatar condition. The \textit{Premade} condition led to a significant over-representation of conversations about the system itself ('META-AI').}
  \label{fig:content_bins}
\end{figure}

We next explored the mechanism behind the avatar effect on rapport. A chi-square test revealed that the thematic content of conversations differed significantly by Avatar Type, $\chi^2(8, N = 162) = 17.47, p = .026$. Standardized residuals showed that participants in the \textbf{Premade} condition were significantly more likely to engage in conversations about the AI system itself (classified as \textbf{META-AI}). In contrast, the creative agency of the \textbf{Generated} condition appeared to inoculate against this detached, evaluative stance (Figure \ref{fig:content_bins}). This suggests a powerful mechanism rooted in psychological ownership and the IKEA effect: by investing their own creative labor, participants in the Generated condition may have transitioned from the role of a system evaluator to that of a co-creator, fostering a more relational, rather than transactional, conversational frame.

\begin{figure}[htbp]
  \centering
  \includegraphics[width=0.8\linewidth]{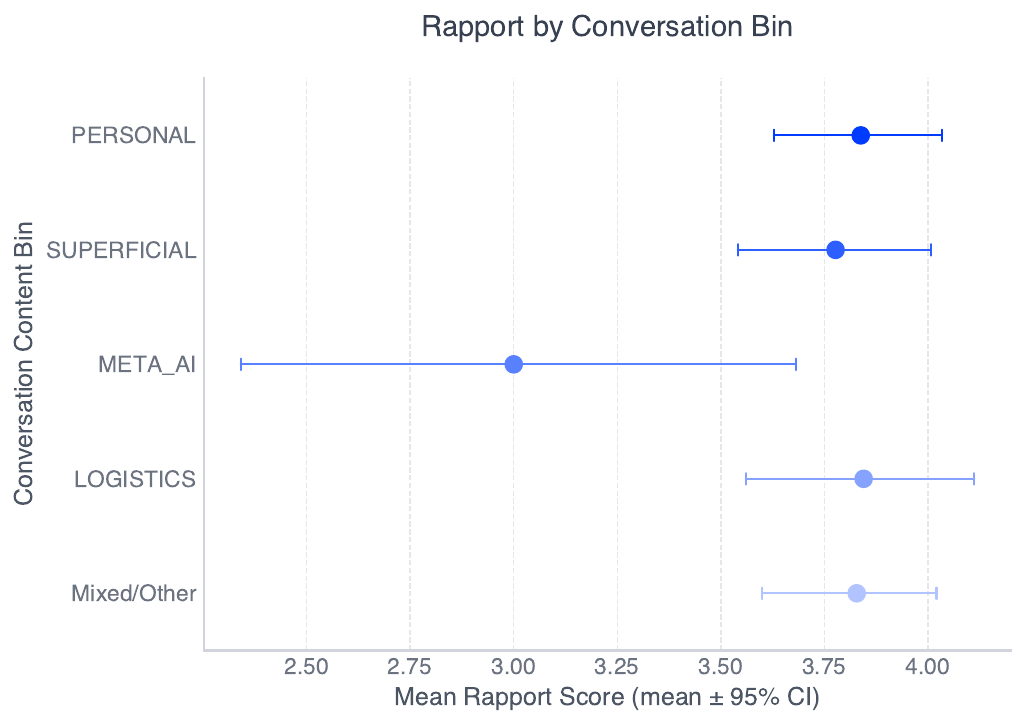}
  \caption{Rapport scores by conversation content bin. Conversations about the AI system ('META-AI') were associated with the lowest rapport, a frame that the \textit{Generated} avatar condition helped users avoid.}
  \label{fig:rapport_by_bin}
\end{figure}

Crucially, conversation content was linked to relational outcomes. An exploratory ANOVA suggested a significant main effect of content bin on rapport ($F(4, 157) = 2.98, p = .021$), with conversations categorized as META-AI associated with the lowest levels (Figure \ref{fig:rapport_by_bin}). However, this effect did not survive a Benjamini-Hochberg FDR correction for multiple comparisons across all seven dependent variables ($p_{\text{adj}} = .073$). This suggests a behavioral pathway where the act of co-creating an avatar reframes the interaction as a social encounter rather than a system evaluation, steering conversation away from detached critique and toward the personal topics that build rapport \citep{collins1994}. Such a shift is particularly potent in human-AI contexts, as users report feeling less judged by and are more willing to disclose sensitive personal information to agents they believe are automated \citep{lucas2014, kumar2025, wang2025}. This effect is particularly pronounced with disembodied agents; increasing an agent's physical, social presence can, conversely, reintroduce social pressures and reduce disclosure of sensitive information \citep{kiesler2008}. This greater willingness to self-disclose, often heightened in interactions with agents perceived as non-judgmental computers \citep{lucas2014, kumar2025}, can in turn establish the sense of \textbf{social presence} that serves as a critical mediator for developing higher-order perceptions like brand partner quality and, ultimately, behavioral intentions \citep{lin2024}. By encouraging more personal conversation, user-generated avatars may have more effectively cultivated this initial sense of social presence.

\FloatBarrier

\section{Discussion}

\subsection{Summary of Findings and the Adaptation Paradox}
This work examined two levers for building rapport with a companion chatbot: user-visible creative agency (avatar generation) and invisible linguistic mimicry (adaptive LSM). In a preregistered 3$\times$2 study, granting users the creative agency to generate their own avatar significantly increased rapport compared to selecting a premade option. Conversely, an algorithmically adaptive linguistic style underperformed a static baseline on perceived personalization and satisfaction. Exploratory analyses suggest a mechanism for this \textbf{Adaptation Paradox}: while the adaptive algorithm was objectively better at maintaining linguistic synchrony, its turn-to-turn variability may have reduced the chatbot's perceived coherence. This aligns with recent findings on expectancy violations in chatbots, where an incongruity between anthropomorphic visual cues (e.g., a humanlike face) and conversational cues (e.g., a mechanical style) leads to negative user perceptions \citep{chen2024, tsai2024}. In short, an agent can be more adaptive in the logs but feel less personal in the user's experience.

For the avatar effect, our analysis points to a specific behavioral mechanism: the act of co-creation appears to foster psychological ownership, which reframes the interaction from a system evaluation to a social encounter, thereby encouraging the more personal conversation that builds rapport. We can model this proposed causal pathway as: \textit{Generated Avatar $\rightarrow$ Psychological Ownership $\rightarrow$ Social Frame $\rightarrow$ Personal Topics $\rightarrow$ Rapport}.

\subsection{Implications for Theory}
These findings contribute to and refine several theoretical streams. First, they nuance anthropomorphism accounts like CASA/MAIN by separating the presence of social cues from their interpretation \citep{nass2000, sundar2008}. User-visible, authored embodiment appears to recruit psychological ownership and the Proteus effect, framing the interaction as a social encounter and encouraging the self-disclosure that builds rapport \citep{pierce2001, yee2007}. This failure can be conceptualized as a form of overaccommodation, where a speaker's attempt to converge overshoots what is necessary or desired, leading to negative evaluations \citep{gasiorek2016}.

Second, our results highlight a critical boundary condition for translating human-human LSM research into human-AI interaction. In human dyads, linguistic synchrony co-varies with shared context and an assumed-coherent persona \citep{ireland2011, giles1991}. In chatbots, synchrony can be engineered without these supports, either by explicitly designing linguistic cues based on predicted user personality \citep{fernau2022}, learning style implicitly from dialogue history \citep{qian2021}, or by rewriting generic responses to conform to a target persona \citep{fu2021}. Our data suggest that stability and legibility are crucial mediators. This can be understood through the distinction between dynamic, turn-by-turn \textit{accommodation} and the perception of a stable, overall \textit{similarity} \citep{postmus2023}. Objective synchrony (i.e., accommodation) is therefore an insufficient design metric on its own, a conclusion supported by sociolinguistic findings that user satisfaction depends on the agent's adherence to a linguistic register appropriate for its social role and context \citep{wang2008, chaves2022}.

Third, the findings refine the adaptability/adaptivity distinction in HCI \citep{schlimbach2023, jones2024}. Adaptability (user-directed change) delivers social benefits via autonomy and authorship. Adaptivity (system-directed change) is not inherently beneficial; its effects are deeply intertwined with other design dimensions, and in some cases, a complementary or non-matched adaptive strategy can outperform one based on simple mimicry \citep{ryong2024}. This tension is evident in practice, where designers must choose between user-driven, autonomy-supportive styles (a form of adaptability) and system-driven, directive styles (a form of adaptivity), with the optimal choice depending on user traits and context \citep{leeuwis2022}. This reframes the design question from ``how much does the system personalize?'' to ``how is personalization surfaced and attributed by the user?''

Our work's central finding, the Adaptation Paradox, offers a clear resolution to the theoretical tension outlined in our background: the results align strongly with Expectancy Violation Theory (EVT) \citep{white2021}. The adaptive linguistic style, while technically synchronous, appears to have created perceptual instability that violated users' expectations of a coherent conversational partner. This finding is consistent with a growing body of recent experimental work demonstrating that a mismatch between a chatbot's visual and conversational cues leads to negative outcomes \citep{chen2024, huang2024, tsai2024}. Specifically, a humanlike visual avatar sets high expectations for conversational fluency; when the agent fails to meet this standard, the visual cue can harm user perceptions \citep{tsai2024}. This underscores the importance of a coherent persona for building trust, as research shows that an AI's perceived integrity—its adherence to a set of understandable principles—is a key antecedent of user trust \citep{lalot2025}. Intriguingly, such expectancy violations are not always detrimental; in some contexts, counter-stereotypical designs—such as a female agent exhibiting stereotypically male competence—can actually increase perceived credibility by productively violating user expectations \citep{schillaci2024}. In our study, however, the adaptive style's variability appears to have created a negative expectation-behavior gap, making a stable, predictable persona feel more "personal" and appropriate.

Finally, the Adaptation Paradox underscores a core requirement for successful parasocial relationships: the persona must be stable and predictable \citep{horton1956}. Users form bonds with a consistent character, a phenomenon now well-documented in studies of social chatbots like Replika, where users develop long-term relationships that can lead to psychological dependence \citep{xie2023, pentina2023}. The perceived instability of our \textit{Adaptive} agent may have felt like it was constantly ``breaking character,'' violating the implicit contract of the parasocial bond, which relies on a stable and predictable persona \citep{horton1956}. This aligns with qualitative analyses of chatbot users who report frustration and a sense of loss when an AI's behavior becomes incoherent or changes unexpectedly after an update \citep{kherraz2024, kettle2024, ly2017, pentina2023}.

\subsection{Implications for Design}
Our findings translate into actionable principles for designing relational AI, aligning with and extending established guidelines for human-AI interaction \citep{amershi2019}. This work complements research in other domains like intelligent tutoring systems, which also highlights the importance of balancing human-like features to foster warmth and engagement without creating user discomfort \citep{nye2014, d2013, wu2024}.

\begin{enumerate}
    \item \textbf{Prioritize User-Directed, Visible Personalization.} Low-friction authorship tasks like avatar generation can yield reliable rapport gains by anchoring the agent’s identity and fostering psychological ownership.
    \item \textbf{Stabilize System-Driven Adaptation.} When implementing linguistic adaptation, designers should impose constraints—such as capping the magnitude of style shifts or smoothing adjustments over several turns—to preserve a coherent persona rather than engaging in raw, volatile mimicry.
    \item \textbf{Make Adaptation Legible.} Instead of relying on covert adaptation, systems should use lightweight cues (e.g., a subtle status indicator, an optional style ``dial'') to help users form an accurate mental model of why and how the agent is adapting.
    \item \textbf{Guard Against Maladaptive Mimicry.} Filter transient user quirks like typos or unusual slang that would violate the agent’s established role. An effective system must distinguish stylistic patterns worth matching from noise. Furthermore, mimicry can fail on a technical level if the underlying models are biased; for instance, speech recognition systems often have higher error rates for speakers of non-standard linguistic varieties, creating a risk that an AI may be unable to accurately perceive or reproduce a user's style \citep{zellou2024}.
\end{enumerate}

\subsection{Ethical and Sociotechnical Considerations}
The ethical tension in adaptive systems is not merely whether to adapt, but how to do so without affective deception. Covert mimicry that is imperceptible to the user risks simulating care while undermining their ability to form an appropriate and calibrated sense of trust, a foundational concern in automation ethics \citep{lee2004, amershi2019}. While human-like cues can increase trust, the effect is highly context-dependent and not always beneficial, highlighting the risk of miscalibration when design is not transparent \citep{klein2025}. Two principles follow. First, \textbf{transparency}: designers should reveal what the system is doing at a granularity that aids sensemaking, as providing explanations has been shown to improve mental models and user trust \citep{lim2009, riveiro2021}. Second, \textbf{agency}: users should be given controls to set persona boundaries (e.g., tone, topics to avoid) and opt-out of mimicry. Given that companion chatbots often address loneliness and mental health, designers must be cautious. While some studies show benefits in reducing loneliness \citep{defreitas2025}, others find heavy usage can correlate with increased dependency and social withdrawal \citep{fang2025}. Our findings suggest that empowering, legible design is not only more effective but also more ethically sound than opaque mimicry.

\subsection{Limitations}
This study has several limitations. First, it captures a single, brief interaction within an open-ended, relational context; as systematic reviews show, findings may differ in the task-oriented contexts that comprise the majority of chatbot research \citep{rapp2021, van2020}. Effects may also differ in longitudinal studies, in mobile contexts, or across cultures with different norms for self-disclosure. Second, our avatar manipulation used static, non-human images; other embodiments (e.g., voice, animation) could alter the effects of agency. Third, our outcomes relied primarily on self-report scales. While we triangulated with behavioral data, future work could benefit from richer in-situ measures. Fourth, our mechanism analyses are exploratory and correlational; the proposed causal paths linking stability to satisfaction and content to rapport require direct experimental manipulation in future work. While our data support a strong hypothesis regarding perceptual coherence, we did not test this as a formal causal model. Finally, for privacy, we release only de-identified derived data, which limits qualitative richness but aligns with our ethical commitments.

\subsection{Future Work}
Three research paths are immediate. First are causal tests of stability and legibility, where smoothing windows or visibility cues are directly manipulated. Second is the exploration of user-facing control surfaces, such as adjustable ``style dials'' for the chatbot's persona. Third are longitudinal field evaluations to examine whether the effects of avatar authorship persist and whether stabilized adaptation improves long-term engagement and well-being.
\FloatBarrier

\section{Conclusion}
This study examined two routes to making companion chatbots feel personal: user-visible authorship through avatar generation and invisible mimicry through adaptive LSM. In a preregistered 3$\times$2 experiment, we found that granting users the creative agency to generate their own avatar reliably increased rapport. Conversely, a technically adaptive linguistic style underperformed a static baseline on key subjective outcomes like perceived personalization and satisfaction. Exploratory analyses suggest this divergence, which we term the \textbf{Adaptation Paradox}, occurs because covert adaptation can reduce a chatbot's perceived coherence, even while maintaining objective linguistic synchrony. An agent can be more adaptive in the logs yet feel less personal in the user's experience.

For HCI, the implication is clear: personalization must be legible, stable, and attributable to a coherent persona to be experienced as care. Our contribution is an empirical demonstration that in relational contexts, visible user agency can outperform covert mimicry. This reframes the evaluation of adaptive systems beyond a simple mantra of ``more synchrony is better'' to a more nuanced consideration of perceptual coherence. Practically, our findings advise designers to prioritize low-friction authorship tools and, when implementing linguistic adaptation, to stabilize and signal those changes to the user.

Future work should causally test the roles of stability and legibility, explore user-facing control surfaces for personalization, and extend these findings to longitudinal and cross-cultural contexts. Ultimately, if there is one insight to carry forward, it is this: personalization only matters if it is seen. The most powerful AI companions will not be those that mirror us invisibly, but those that empower us to co-author the interaction—and, in doing so, reflect back something we recognize as meaningfully human.

\noindent\textit{Preprint note (Code availability):} Kagami platform code used in this study is archived at \href{https://doi.org/10.5281/zenodo.15801081}{https://doi.org/10.5281/zenodo.15801081}.

\nocite{brandt2025}
\clearpage
\appendix

\section{Survey Instruments}
\label{sec:appendix_survey}
This section contains all items from the pre-interaction and post-interaction surveys. Each item is mapped to its corresponding variable name.

\subsection*{Pre-Survey Instrument}

\subsubsection*{1. Demographics}
\begin{enumerate}
    \item \textbf{What is your age?} (Variable: \texttt{Demo1})
    \item \textbf{What is your gender identity?} (Variable: \texttt{Demo2})
    \item \textbf{Which of the following best describes your race/ethnicity? (Select all that apply.)} (Variable: \texttt{Demo3})
    \item \textbf{What is the highest level of education you have completed?} (Variable: \texttt{Demo4})
\end{enumerate}

\subsubsection*{2. Attitudes and Personality (AP1 Block)}
\textit{Response scale: 5-point Likert (1 = Strongly disagree, 5 = Strongly agree).}
\begin{enumerate}
    \item I generally enjoy exploring and using new technologies. (Variable: \texttt{AP1\_1})
    \item In general, I believe AI chatbots can be useful. (Variable: \texttt{AP1\_2})
    \item I feel uncomfortable when interacting with AI systems. (Variable: \texttt{AP1\_3})
    \item I am generally skeptical about AI systems. (Variable: \texttt{AP1\_4})
    \item AI systems make me feel uneasy. (Variable: \texttt{AP1\_5})
    \item I see myself as someone who is open to new experiences. (Variable: \texttt{AP1\_6})
    \item I see myself as someone who is outgoing and sociable. (Variable: \texttt{AP1\_7})
    \item I see myself as someone who is methodical and organized. (Variable: \texttt{AP1\_8})
    \item I see myself as someone who is calm and emotionally stable. (Variable: \texttt{AP1\_9})
    \item I see myself as someone who is sympathetic and warm. (Variable: \texttt{AP1\_10})
    \item Most people are trustworthy. (Variable: \texttt{AP1\_11})
\end{enumerate}

\subsubsection*{3. Loneliness (ULS-4)}
\textit{Adapted from \citet{russell1978}. Response scale: 5-point Frequency (1 = Never, 5 = Always).}
\begin{enumerate}
    \item How often do you feel that you lack companionship? (Variable: \texttt{WB1\_1})
    \item How often do you feel left out? (Variable: \texttt{WB1\_2})
\end{enumerate}

\subsubsection*{4. Depression (PHQ-2)}
\textit{Items from \citet{kroenke2003}. Response scale: 4-point Frequency (1 = Not at all, 4 = Nearly every day).}
\begin{enumerate}
    \item Little interest or pleasure in doing things. (Variable: \texttt{WB2\_1})
    \item Feeling down, depressed, or hopeless. (Variable: \texttt{WB2\_2})
\end{enumerate}

\subsubsection*{5. Anxiety (GAD-2)}
\textit{Items from \citet{spitzer2006}. Response scale: 4-point Frequency (1 = Not at all, 4 = Nearly every day).}
\begin{enumerate}
    \item Feeling nervous, anxious, or on edge. (Variable: \texttt{WB2\_3})
    \item Not being able to stop or control worrying. (Variable: \texttt{WB2\_4})
\end{enumerate}

\subsection*{Post-Survey Instrument}
\subsubsection*{1. Manipulation Checks}
\begin{enumerate}
    \item \textbf{Which of the following best describes the chatbot's appearance during your conversation?} (Variable: \texttt{MC1})
    
    \item \textbf{(If avatar was present) Please rate your experience with the chatbot's appearance.} \textit{(5-point Likert scale)}
    \begin{enumerate}[label=\alph*.]
        \item The chatbot's avatar felt like an important part of the interaction. (Variable: \texttt{MC2\_1})
        \item I was satisfied with the appearance of the chatbot's avatar. (Variable: \texttt{MC2\_2})
        \item I felt I was able to choose or customize the chatbot's appearance. (Variable: \texttt{MC2\_3})
        \item The chatbot's avatar looked similar to how I see myself. (Variable: \texttt{MC2\_4})
    \end{enumerate}
    
    \item \textbf{(If avatar was present) How lifelike did the chatbot's avatar feel to you?} (Variable: \texttt{MC3}) \textit{(5-point scale: 1 = Not at all lifelike, 5 = Extremely lifelike)}
    
    \item \textbf{Please rate the chatbot's communication style.} \textit{(5-point Likert scale)}
    \begin{enumerate}[label=\alph*.]
        \item The chatbot's language style seemed to adapt to the way I was talking. (Variable: \texttt{MC4\_1})
        \item The chatbot's way of talking felt personalized to me. (Variable: \texttt{MC4\_2})
        \item The chatbot used a consistent language style throughout the conversation. (Variable: \texttt{MC4\_3})
    \end{enumerate}
    
    \item \textbf{(If no avatar was present) I think the conversation would have felt more engaging if the chatbot had an avatar.} (Variable: \texttt{MC5}) \textit{(5-point Likert scale)}
\end{enumerate}

\subsubsection*{2. Core Dependent Variables}
\textit{All items rated on a 5-point Likert scale (1 = Strongly disagree, 5 = Strongly agree).}

\begin{description}
    \item[Anthropomorphism] \textit{(Adapted from \citet{blut2021}, \citet{go2019}, \citet{wu2024}.)}
        \begin{enumerate}
            \item The chatbot displayed characteristics typical of a living being. (Variable: \texttt{CDV1\_1})
            \item The chatbot seemed to have a distinct personality. (Variable: \texttt{CDV1\_2})
            \item Interacting with the chatbot felt like interacting with another social being. (Variable: \texttt{CDV1\_3})
        \end{enumerate}
        
    \item[Perceived Personalization] \textit{(Adapted from \citet{paul2024}, \citet{wu2024}, \citet{sidlauskiene2023}.)}
        \begin{enumerate}
            \item The chatbot felt tailored to me personally. (Variable: \texttt{CDV1\_4})
            \item The chatbot's responses felt personalized. (Variable: \texttt{CDV1\_5})
            \item The chatbot seemed to understand my preferences. (Variable: \texttt{CDV1\_6})
            \item I felt like the chatbot was interacting specifically with me. (Variable: \texttt{CDV1\_7})
        \end{enumerate}
        
    \item[Trust] \textit{(Adapted from \citet{mayer1995}, \citet{go2019}.)}
        \begin{enumerate}
            \item I felt I could trust the chatbot. (Variable: \texttt{CDV1\_8})
            \item The chatbot seemed reliable. (Variable: \texttt{CDV1\_9})
            \item The chatbot was capable. (Variable: \texttt{CDV1\_10})
            \item The chatbot seemed to have my best interests at heart. (Variable: \texttt{CDV1\_11})
            \item The chatbot seemed honest. (Variable: \texttt{CDV1\_12})
        \end{enumerate}
        
    \item[Rapport] \textit{(Adapted from \citet{vossen2024}, \citet{hoffman2023}, \citet{colton2024}.)}
        \begin{enumerate}
            \item I felt a strong connection with the chatbot. (Variable: \texttt{CDV1\_13})
            \item I felt comfortable interacting with the chatbot. (Variable: \texttt{CDV1\_14})
            \item I felt like the chatbot cared about what I had to say. (Variable: \texttt{CDV1\_15})
            \item I felt respected by the chatbot. (Variable: \texttt{CDV1\_16})
        \end{enumerate}
        
    \item[Social Presence] \textit{(Adapted from \citet{biocca2003}, \citet{go2019}.)}
        \begin{enumerate}
            \item I felt a sense of presence from the chatbot. (Variable: \texttt{CDV1\_17})
            \item I felt like there was a ``person'' on the other side. (Variable: \texttt{CDV1\_18})
            \item The chatbot felt real to me. (Variable: \texttt{CDV1\_19})
            \item The chatbot's avatar or appearance helped create a sense of presence. (Variable: \texttt{CDV1\_20})
        \end{enumerate}
        
    \item[Engagement] \textit{(Adapted from \citet{xie2023}, \citet{hoffman2023}.)}
        \begin{enumerate}
            \item I enjoyed interacting with the chatbot. (Variable: \texttt{CDV1\_21})
            \item Interacting was fun. (Variable: \texttt{CDV1\_22})
            \item I felt absorbed and engaged in the conversation. (Variable: \texttt{CDV1\_23})
            \item I would like to interact with the chatbot again. (Variable: \texttt{CDV1\_24})
        \end{enumerate}
        
    \item[Satisfaction] \textit{(Adapted from \citet{janson2023}, \citet{paul2024}, \citet{blut2021}.)}
        \begin{enumerate}
            \item Overall, I am satisfied with my experience. (Variable: \texttt{CDV1\_25})
            \item The quality of responses was satisfactory. (Variable: \texttt{CDV1\_26})
            \item The interaction was valuable. (Variable: \texttt{CDV1\_27})
            \item The chatbot was helpful. (Variable: \texttt{CDV1\_28})
        \end{enumerate}
\end{description}

\subsubsection*{3. Open-Ended Feedback}
\begin{enumerate}
    \item \textbf{What did you like most about your interaction with Kagami?} (Variable: \texttt{OE1})
    \item \textbf{What did you like least about your interaction with Kagami?} (Variable: \texttt{OE2})
    \item \textbf{(Optional) How did the avatar's appearance influence your experience?} (Variable: \texttt{OE3})
    \item \textbf{(Optional) Did you notice anything about the chatbot's communication style? If so, what?} (Variable: \texttt{OE4})
\end{enumerate}

\section{Interface Stimuli}
\label{sec:appendix_stimuli}

This appendix contains images of the user interface for the experimental chatbot platform. These visuals illustrate the key steps of the user flow and the manifestation of the experimental conditions described in the main paper.

\subsection*{Disclaimer and Setup Screens}
All participants first saw a disclaimer screen. Those in the avatar conditions then proceeded to an avatar selection or generation screen before beginning the chat.

\begin{figure}[htbp]
    \centering
    \includegraphics[width=0.5\linewidth]{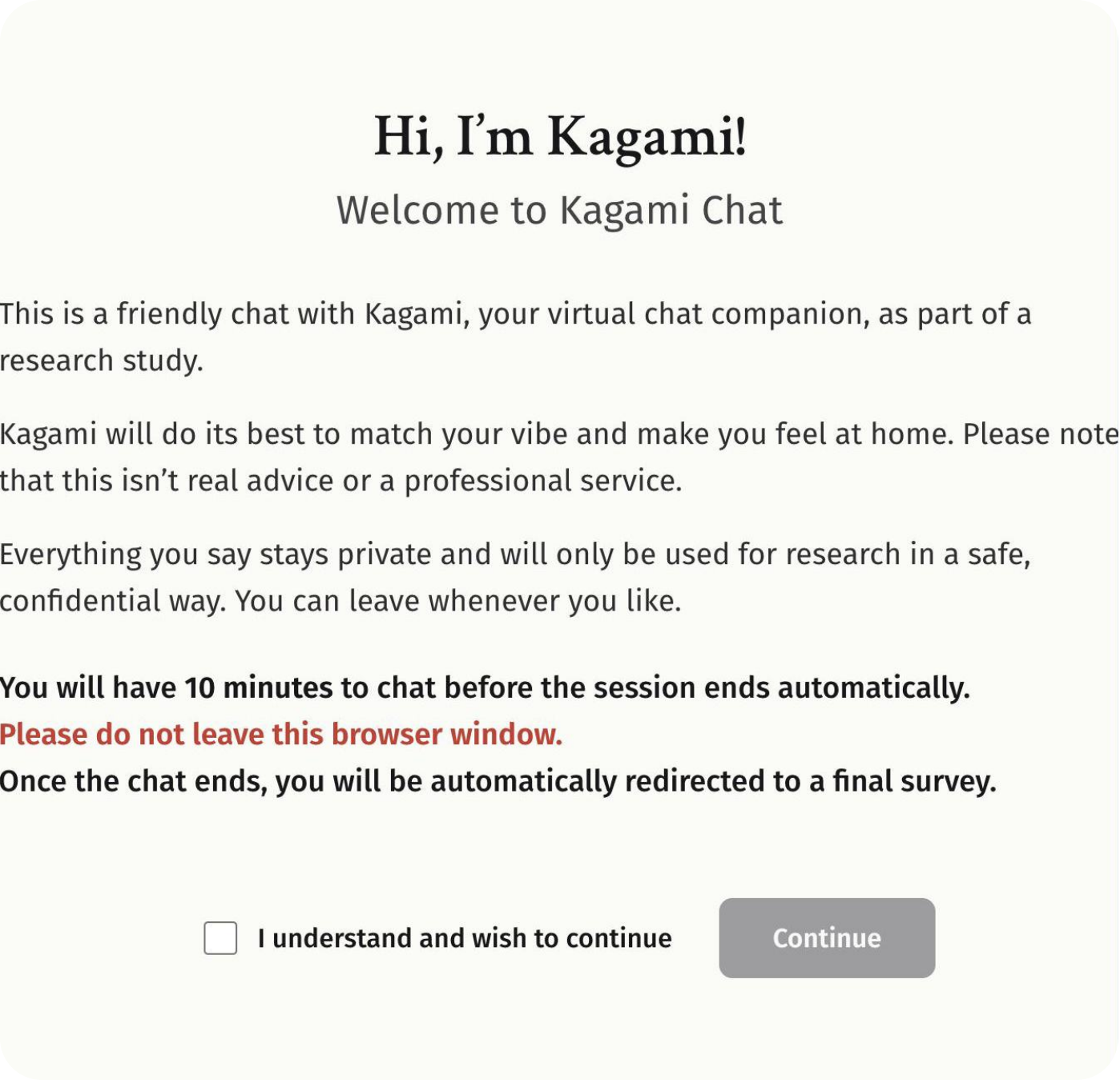}
    \caption{Disclaimer screen shown to participants in the `none` condition.}
    \label{fig:app_disclaim_none}
\end{figure}

\begin{figure}[htbp]
    \centering
    \includegraphics[width=0.5\linewidth]{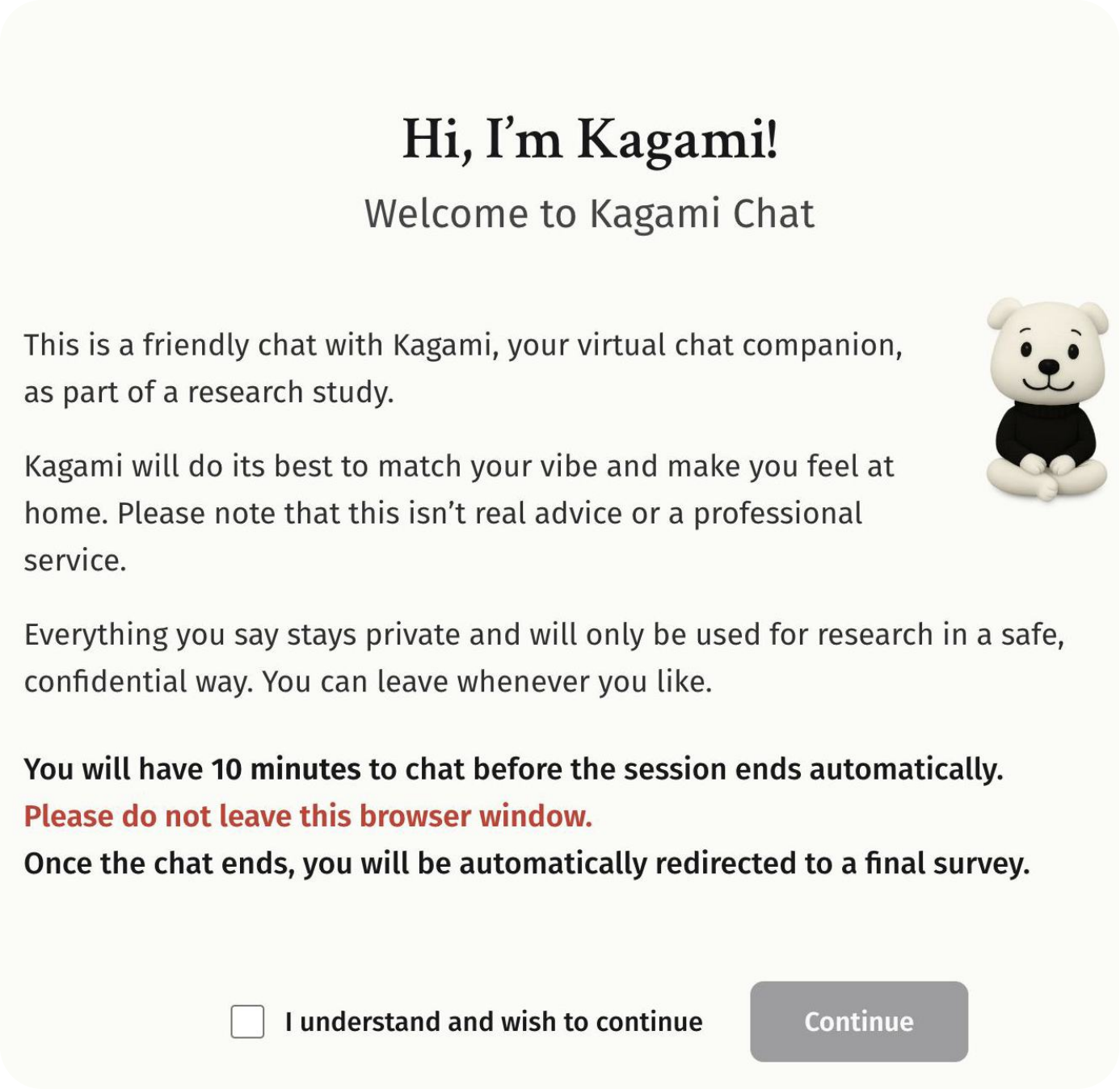}
    \caption{Disclaimer screen shown to participants in the `premade` and `generated` avatar conditions.}
    \label{fig:app_disclaim_avatar}
\end{figure}

\begin{figure}[htbp]
    \centering
    \includegraphics[width=0.5\linewidth]{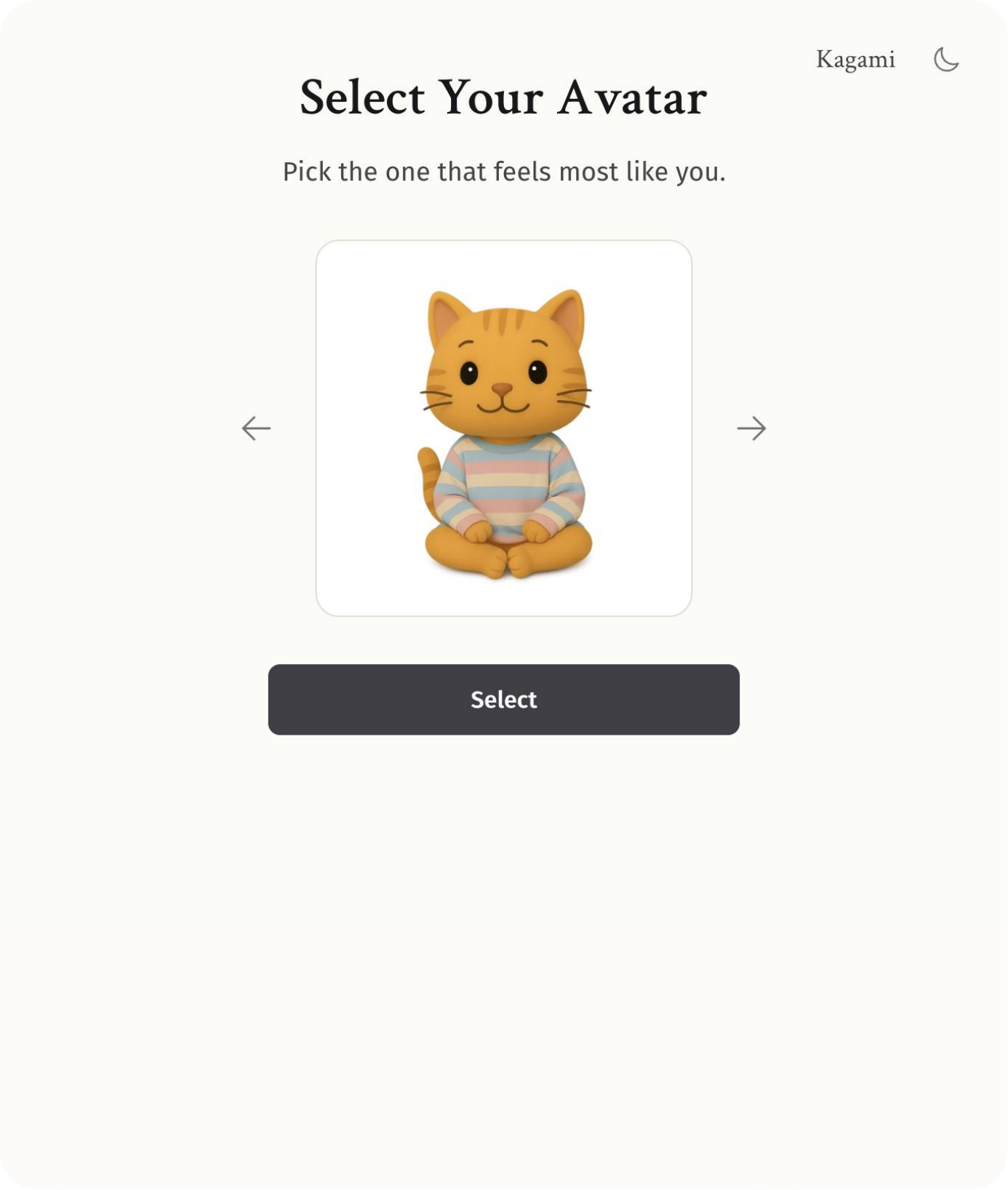}
    \caption{The premade avatar selection screen. Participants in this condition chose an avatar from a carousel.}
    \label{fig:app_avatar_premade}
\end{figure}

\begin{figure}[htbp]
    \centering
    \includegraphics[width=0.5\linewidth]{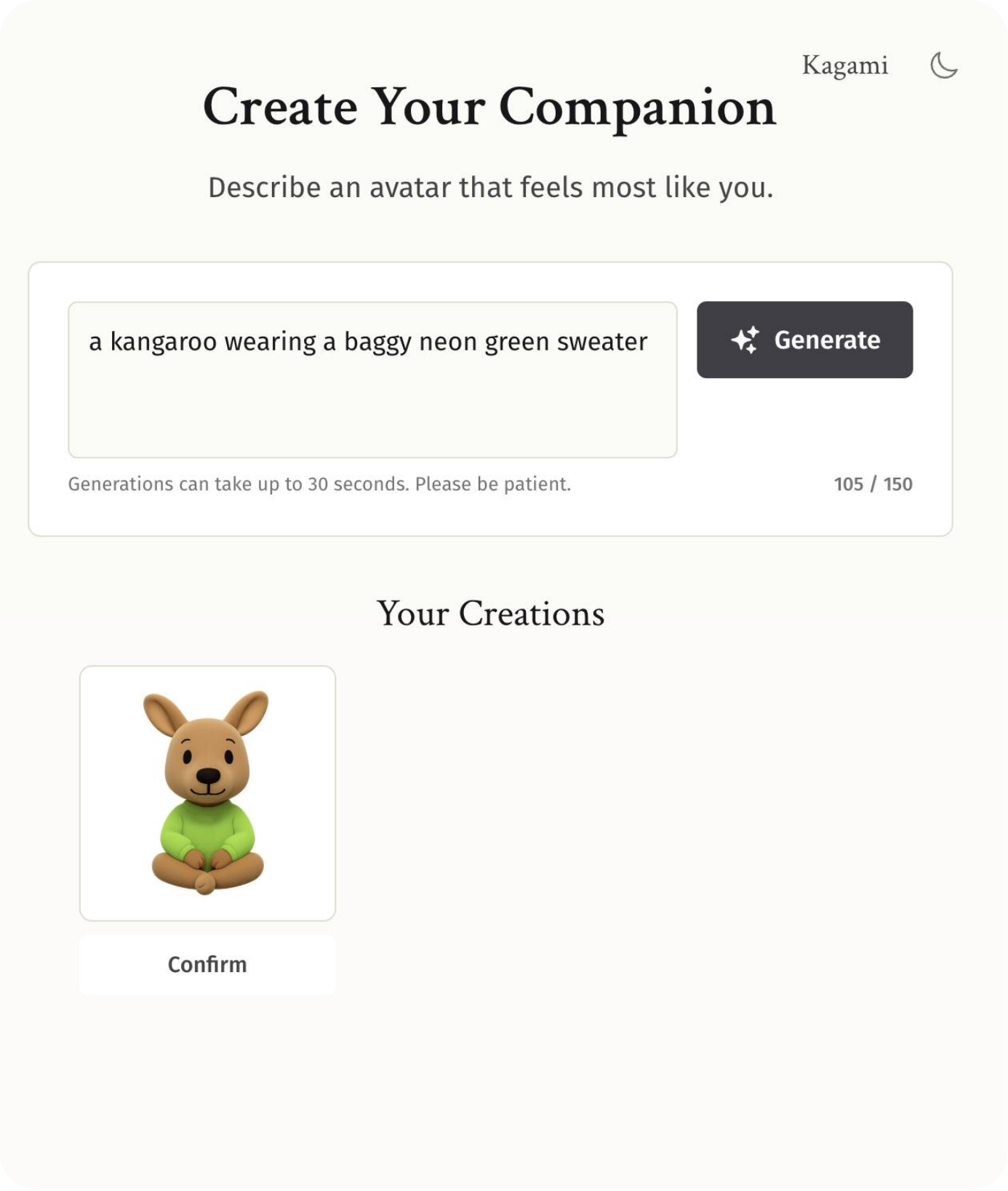}
    \caption{The generated avatar creation screen. Participants in this condition wrote a text prompt to create a custom avatar.}
    \label{fig:app_avatar_generate}
\end{figure}

\subsection*{Resulting Chat Screens}
After the setup phase, participants began their 10-minute timed conversation. The chat interface visually reflected the assigned `Avatar Type` condition.

\begin{figure}[htbp]
    \centering
    \includegraphics[width=0.5\linewidth]{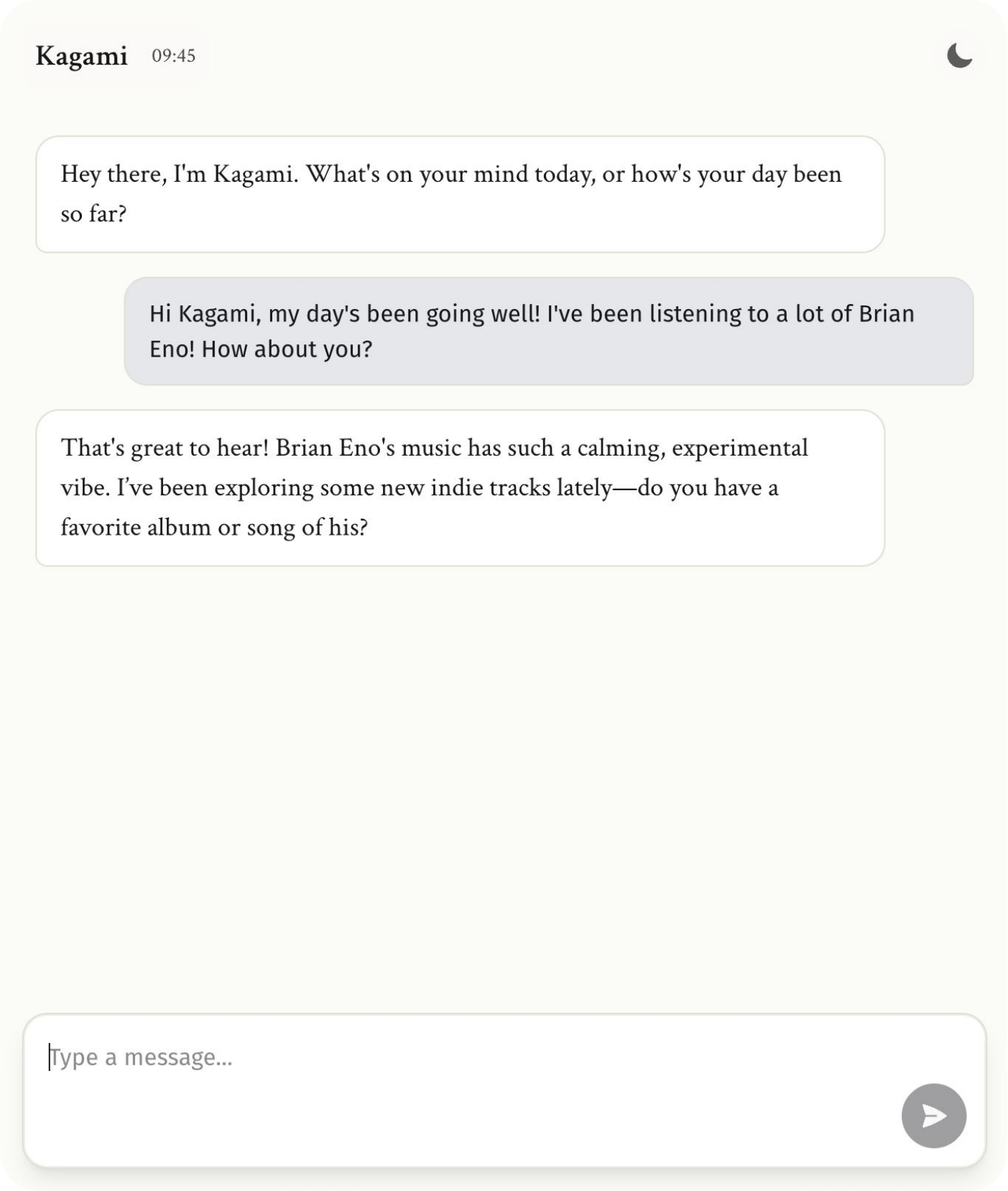}
    \caption{Chat screen for the `none` condition. No avatar representation was shown.}
    \label{fig:app_chat_none}
\end{figure}

\begin{figure}[htbp]
    \centering
    \includegraphics[width=0.5\linewidth]{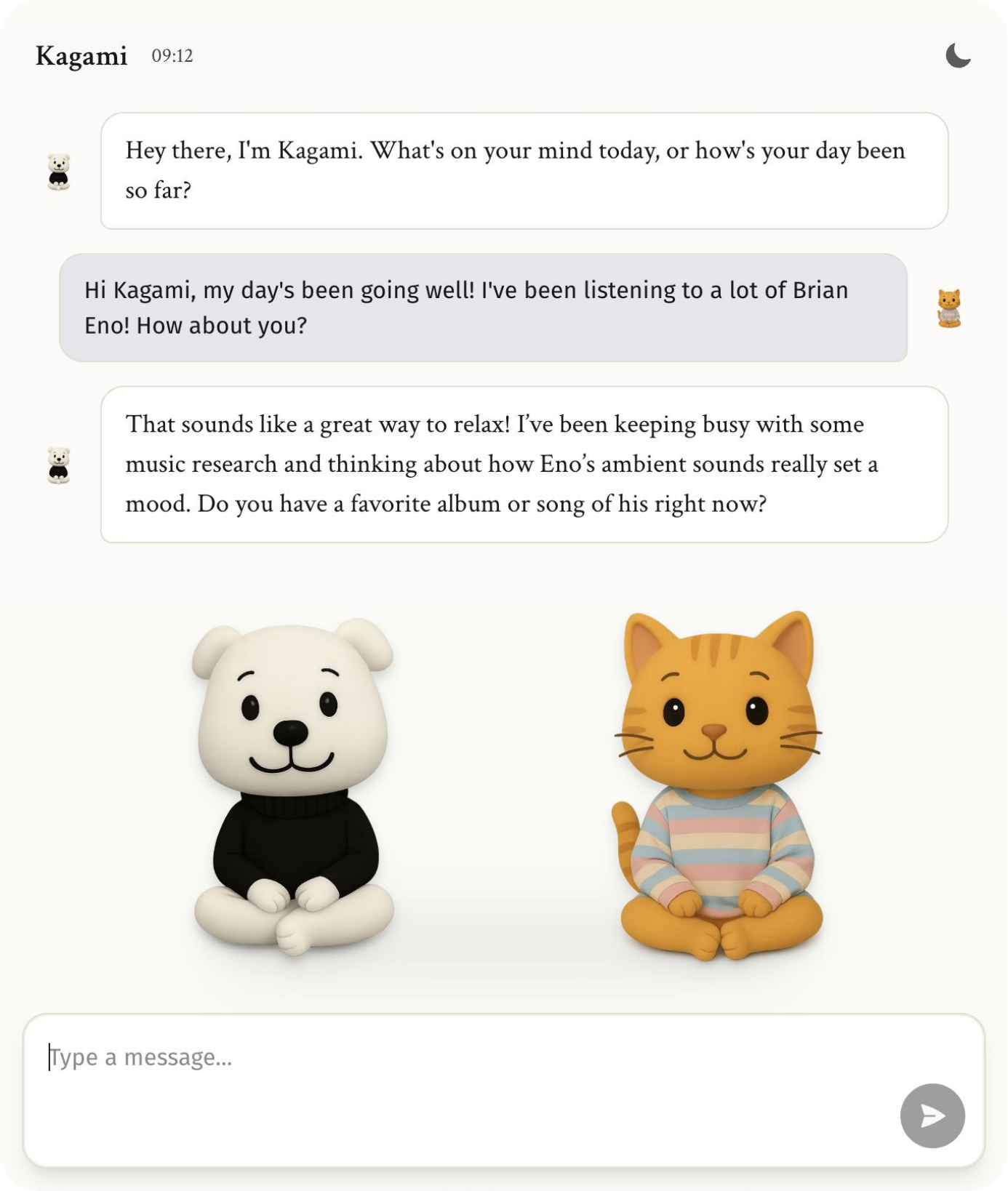}
    \caption{Chat screen for the `premade` condition. The user’s selected avatar was displayed.}
    \label{fig:app_chat_premade}
\end{figure}

\begin{figure}[htbp]
    \centering
    \includegraphics[width=0.5\linewidth]{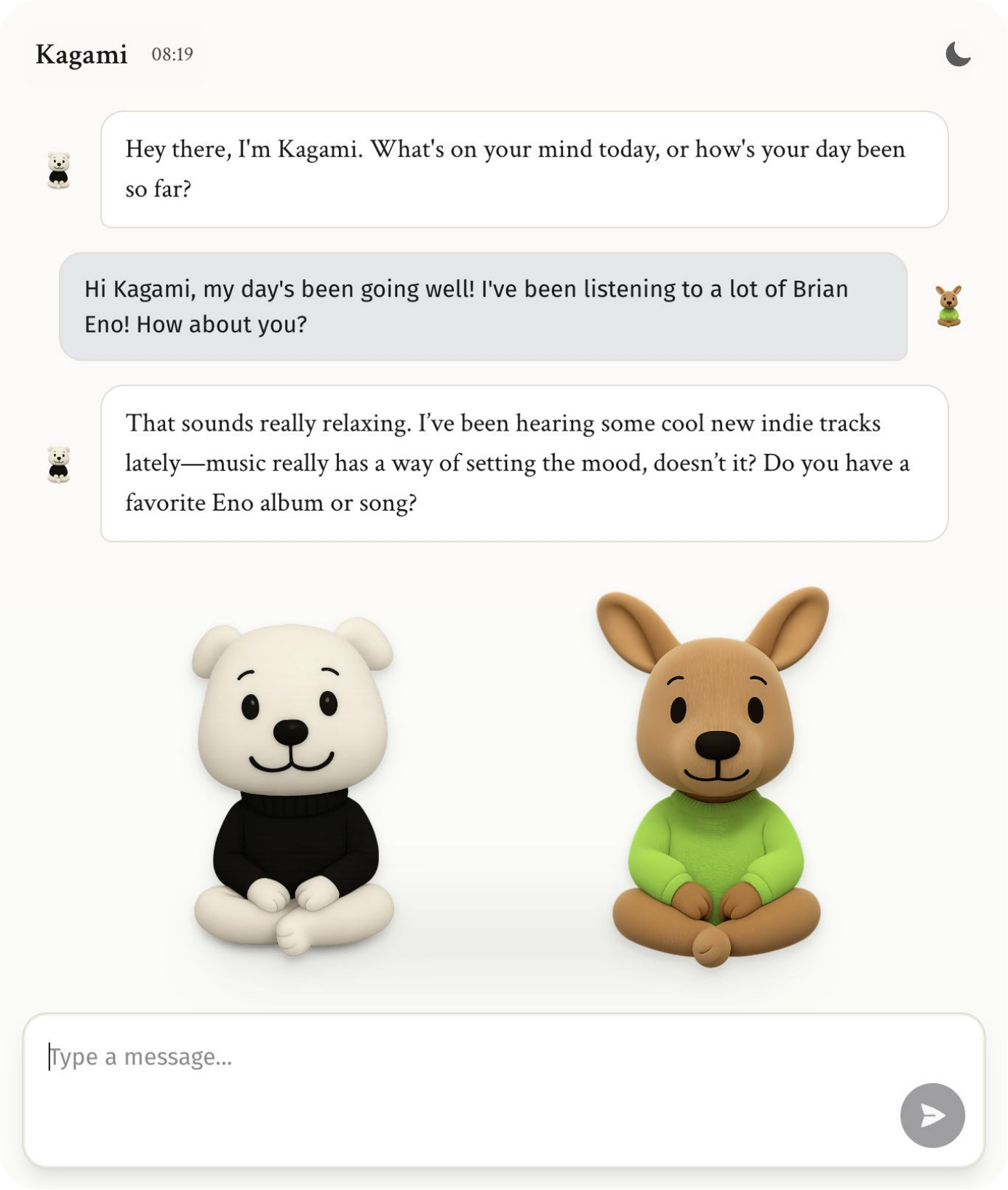}
    \caption{Chat screen for the `generated` condition. The user’s generated avatar was displayed.}
    \label{fig:app_chat_generated}
\end{figure}

\FloatBarrier 

\section{Recruitment and Power Analysis}
\label{sec:appendix_recruitment}

\subsection{Recruitment Materials}
The following text was used for the study listing posted on the Prolific online research platform to recruit participants.

\subsubsection*{Prolific Study Title: Chat with an AI Companion and Share Your Feedback}

\paragraph{Prolific Study Description} In this academic research study, you will have a brief, friendly chat with a companion AI named Kagami and then answer some survey questions about your experience.

\paragraph{What you will do:}
\begin{itemize}
    \item Complete a short background survey (2-3 minutes).
    \item Have a 10-minute, text-based conversation with the AI.
    \item Complete a final survey about your perceptions (3-5 minutes).
    \item The conversation will be about general, everyday topics. There are no right or wrong answers; we are simply interested in your experience. You are required to use a desktop or laptop computer.
\end{itemize}

\paragraph{Payment:}
\begin{itemize}
    \item Reward: \$3.50
    \item Estimated Completion Time: 15-20 minutes
    \item Estimated Hourly Rate: \$10.50 / hour
\end{itemize}

\paragraph{Eligibility Criteria (Used for Prolific Pre-screening):}
\begin{itemize}
    \item Age: 18+ years
    \item Location: Residing in the United States
    \item Language: Fluent in English
    \item Previous Participation: Have not participated in pilot testing for this specific ``Kagami'' study.
\end{itemize}

\subsection{A Priori Power Analysis}
\begin{figure}[h!]
  \centering
  \includegraphics[width=0.8\columnwidth]{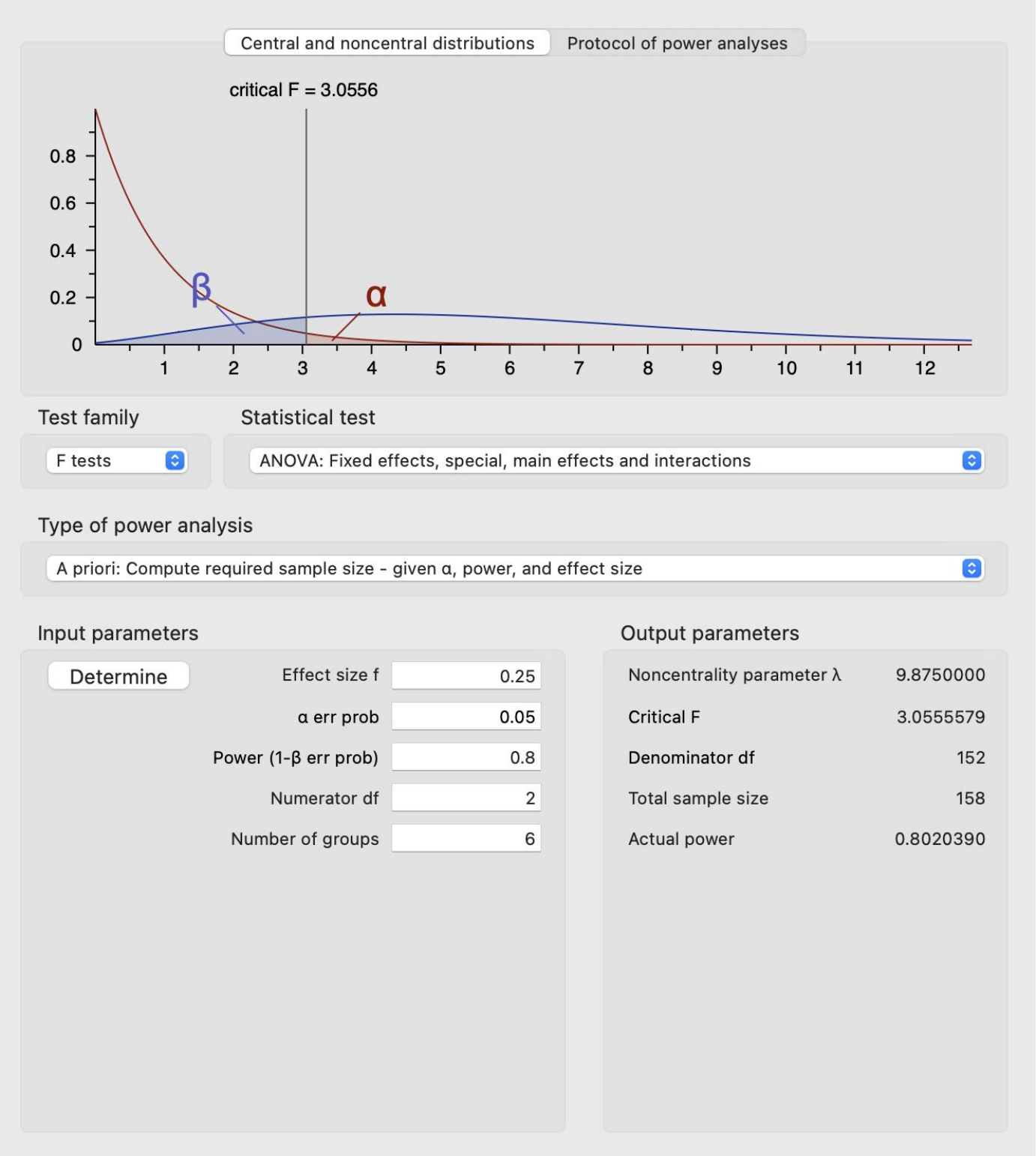}
  \caption{A Priori Power Analysis in G*Power 3.1 for a 3 × 2 ANOVA. The analysis determined a required sample size of 158 to detect a medium effect size (\textit{f} = 0.25) with 80\% power at $\alpha$ = .05.}
  \label{fig:gpower}
\end{figure}

An a priori power analysis was conducted using G*Power 3.1 to determine the required sample size for the study’s 3 × 2 between-subjects ANOVA design \citep{faul2007}. The analysis was based on detecting a medium effect size (\textit{f} = 0.25) with 80\% statistical power at an alpha level of $\alpha$ = .05. Results indicated that a total of 158 participants would be needed. A target sample size of N = 162 was selected to accommodate potential data exclusions (see Figure~\ref{fig:gpower}).

\FloatBarrier 

\clearpage
\bibliography{references}

\end{document}